\documentclass[preprint,12pt]{aastex}
\usepackage{graphicx}
\usepackage{amssymb, amsmath, chemarrow}

\usepackage{natbib}

\newcommand\cc{\mbox{cm$^{-3}$}}
\newcommand\DtwoHp{\mbox{${\rm D_2H^+}$}}
\newcommand\HtwoDp{\mbox{${\rm H_2D^+}$}}
\newcommand\Dthreep{\mbox{${\rm D_3^+}$}}
\newcommand\Hthreep{\mbox{${\rm H_3^+}$}}
\newcommand\Htwo{\mbox{${\rm H_2}$}}
\newcommand\nhtwo{\mbox{${\rm n(H_2)}$}}
\newcommand\HtwoO{\mbox{${\rm H_2O}$}}
\newcommand\Htwoopr{\mbox{${\rm H_2}$ {\em opr}}}
\newcommand{\hdtp}{${\rm HD}_2^+$}
\newcommand{\dt}{D$_{2}$}
\newcommand{\dthp}{D$_{3}^+$}
\newcommand{\htdp}{H$_{2}$D$^+$}
\newcommand{\hh}{H$_{2}$}
\newcommand{\hthp}{H$_{3}^+$}
\newcommand{\ra}{$\rightarrow$}

\begin{document}

\title{The D/H Ratio of Water Ice at Low Temperatures}
\shorttitle{Deuterium in Water Ice}
\shortauthors{Lee \& Bergin}
\author{Jeong-Eun Lee}
\affil{Department of Astronomy and Space Science, Kyung Hee University,
  Yongin-si, Gyeonggi-do 446-701, Korea}
\and
\author{Edwin A. Bergin}
\affil{ Department of Astronomy, The University of Michigan,
500 Church Street, Ann Arbor, Michigan 48109-10424}

\begin{abstract}
We present the modeling results of deuterium fractionation of water ice, H$_2$,  and the primary deuterium isotopologues of \Hthreep\ 
 adopting physical conditions associated with the star  and planet formation process.    We calculated the deuterium chemistry for a range of gas temperatures (T$_{gas} \sim 10 - 30$~K), molecular hydrogen density (n(H$_2)\sim 10^4 - 10^7$), and ortho/para ratio ({\em opr}) of H$_2$ based on state-to-state reaction rates and explore the resulting fractionation including the formation of a water ice mantle coating grain surfaces.  We find that the deuterium fractionation exhibits the expected temperature dependence of large enrichments at low gas temperature.  
 More significantly the inclusion of water ice formation leads to large D/H ratios in water ice ($\gtrsim 10^{-2}$ at 10~K) but also alters the overall deuterium chemistry.   For T $<$ 20~K the implantation of deuterium into ices lowers the overall abundance of HD which reduces the efficiency of deuterium fractionation at high density. In agreement with an earlier study, under these conditions HD may not be the primary deuterium reservoir in the cold dense interstellar medium and \Hthreep\ will be the main charge carrier in the dense centers of pre-stellar cores and the protoplanetary disk midplane.
 \end{abstract}

\section{Introduction}

The fractionation of deuterium is an important chemical tracer of the physical history of the
dense interstellar medium.   The initiating reaction for the chemistry
(${\rm H_3^+ + HD}$ $\rightleftharpoonsfill{2.5em}$ ${\rm H_2D^+ + H_2}$) is
 exothermic in the forward direction, requiring additional energy of $\sim 232$~K to activate 
 the backwards channel.    As such large deuterium enrichments in gaseous species have been 
 isolated at low (T$_{gas} \sim 10$~K)  temperatures  in dense (n$_{H_2}$ $> 10^4$ \cc ) star-forming gas 
\citep[e.g. DCO$^+$, DCN;][]{millar_dfrac, rg03}.  A key facet of the gas-phase chemistry is that many of the deuterium fractionation
reactions produce an excess of deuterium atoms relative to hydrogen atoms.  Due to high mobility, atoms are key players in the surface chemistry of molecular ices.  This leads to the prediction that molecular ices formed at low temperatures will be enriched in deuterium \citep{tielens83}.    Indeed high-levels of D-fractionation of water has long been observed to be present in warm (T $\gtrsim$ 100~K) gas near young massive stars \citep{jacq90}.   At these temperatures the D/H ratio of gaseous molecules should be commensurate with the interstellar D/H ratio of $\sim 3 \times 10^{-5}$ \citep{linsky06} and the D-enriched water is therefore believed to be originated from ices that formed at cold (T$_{gas}$ = T$_{dust}$ $\sim 10-30$~K) temperatures prior to the formation of the star.

Historically the measured water D/H enrichment in massive hot cores ranged from $\sim$ 20 --100 above the galactic value
\citep{jacq90, gmw96, helmich96}.
The enriched D/H ratio of water has been also detected in the warm inner envelopes of low mass YSOs \citep{stark04, parise05}.  
The study of water in the interstellar medium has recently been given a large boost from the successful launch and operations of the {\em Herschel} Space Observatory \citep{pilbratt10}.   {\em Herschel} has observed water vapor in numerous star-forming regions \citep{evd_wish}, which can be supplemented with ground-based or Herschel observations of HDO to derive a D/H ratio.
Some of the {\em Herschel} results are finding high D/H ratios of water.  For instance, \citet{bergin10a} detected HD$^{18}$O in Orion KL which implies a water D/H ratio of $\sim 0.003$ \citep{neill13a}.
Towards some low mass protostars, in both resolved and unresolved data, high levels of enrichments are found with HDO/H$_2$O $>$ 0.003-0.05
\citep{liu11, coutens12, taquet13b}.   However, in some instances lower values are found; for example \citet{jj10} set an upper limit of HDO/H$_2$O $\sim 6 \times 10^{-4}$ in one source while \citet{persson13} estimate $\sim 9 \times 10^{-4}$ in yet another.
Thus, a range of enrichment levels are now being inferred to have been present in cold water ice.  This implied diversity calls for a theoretical exploration of how deuterium is implanted into ices.

Over the past few years  there have been two advances in our theoretical understanding of {\em gas-phase} deuterium fractionation in the interstellar medium that have impact on the composition of both gas and ices.  First, we have recognized that the deuterium fractionation does not stop with H$_2$D$^+$ as the sole initiating ion in the chemistry.  Rather successive reactions with HD can form additional deuterated isotopologues of H$_3^+$:

\begin{displaymath}
\rm {H_2D^+ \autorightleftharpoons{HD}{H$_2$} D_2H^+ \autorightleftharpoons{HD}{H$_2$} D_3^+}
\end{displaymath}

\noindent \citep[each is exothermic in the forward direction with a barrier for the reverse channel;][]{vastel04, roberts_dfrac}.   This has important implications for the primary charge carriers and also the products of deuterium chemistry, which can trend to multiply deuterated forms \citep{roberts_dfrac}.   Because HD and H$_2$ do not freeze in appreciable abundance on grain surfaces some of these molecules (e.g. H$_2$D$^+$ and D$_2$H$^+$) have been isolated as tracers of the gas physics in the heavy-element freeze-out dominated dense centers of pre-stellar cores \citep{flower_dfrac, vdt_l1544}.

  Second, it has been noted that deuterium fractionation reactions have an innate dependence on the ortho-to-para ratio ({\em opr}) of H$_2$ \citep{gdr02} and also on the relative abundances of the high and low energy spin isomers of each of the reactants \citep[H$_2$D$^+$, D$_2$H$^+$, D$_3^+$, H$_3^+$;][]{flower_dfrac, hugo09}.    Thus the presence of even a small amount of ortho-H$_2$, which has excess internal energy, can aid in overcoming the barrier for the reverse channel in the above reactions.  This will reduce fractionation at the start of the chain \citep[][hereafter WFP04]{walmsley_dep}, while the relative spin state abundances of the other reactants can also change the resulting distribution of fractionated products.
  
In our work we wish to explore the deuterium fractionation that becomes implanted within ices
in the context of a cold and dense core and its dependence on the H$_2$ {\em opr}, the gas temperature,
and the gas density.
According to \citet{flower06}, the {\em opr} for key species (H$_2$D$^+$, D$_2$H$^+$, D$_3^+$, H$_3^+$, H$_2$)  vary during the core formation process (i.e., as the density grows).  Furthermore there is an additional dependence on the dynamical evolution (e.g., free-fall or steady-state contraction).  In this work, we do not adopt any dynamical model for the core evolution. We rather consider the physical conditions (density and temperature) as free, but fixed, parameters for a given model.  In this sense a higher density can be considered as representative for later core evolutionary stages. We also consider the H$_2$ {\em opr} as a free parameter since it is 
unlikely in the steady state or equilibrium state (LTE) unless the timescale is long enough (see Fig. 1 of  \citet{flower06}). Finally, in this study, the chemical evolution is calculated at a given physical condition with pre-calculated {\em opr}s. 

In \S 2 we will present our model of the deuterium chemistry, while \S 3 presents the results showing variable levels of water-ice deuterium fractionation can be created depending on the H$_2$ {\em opr} and the gas temperature as well as the H$_2$ density.  This can significantly alter the HD abundance and the composition of major charge carriers in the dense core center. In \S 4 we discuss the implications of our result.

\section{Model}

We adopt the chemical code used in \citet{bergin95}, \citet{bergin97},  and \citet{lee04}. The code deals with the basic gas chemistry 
(photodissociation, photoionization, and cosmic ray ionization) and gas-grain interactions (adsorption and desorption). 
The binding energies of H, O, OH, and H$_2$O are adopted from \citet{ch07}, and they are 650, 800, 3500, and 5640 K, respectively, for the water coated grain surfaces.
For the case of bare silicate grain surfaces, these binding energies are divided by 1.47 \citep{lee04}.  
The sticking coefficient is assumed as 1.0 for all species including H and D.
According to \citet{chaabouni12}, the sticking coefficients of H and D are 0.86 and 0.95, respectively, to a bare silicate grain surface. 
For water ice coated grains, the sticking coefficients must be greater. Therefore, we assume unity for the sticking coefficients of H and D.
Three desorption mechanisms are included: thermal evaporation, cosmic-ray-induced heating, and direct photodesorption. 
We have assumed the H$_2$ cosmic-ray ionization rate ($\zeta$) of $6\times10^{-17}$ s$^{-1}$, which results in the equilibrium number density of atomic hydrogen of ~2.5 \cc\ in our model; 
these numbers are consistent with the result of \citet{gl05}.
We have updated the code to include the ortho-to-para dependent deuterium chemistry in the gas and the surface chemistry for the formation of H$_2$ and water ice as described 
in the subsections below.
Our chemical network includes $\sim 170$ species (including grain counterparts and deuterium counterparts) and $\sim 1200$ reactions. 

Motivated by \citet[][hereafter HHB10]{hhb10} we assume the core is evolving from the lower density
molecular cloud with most hydrogen as molecular, carbon in CO, and some initial water ice mantle present.
In reality, a dense core that we model in this work must have evolved from lower density material. 
Chemical abundances in a forming core also change with time and density. However, most studies of dense core chemical evolution assume atomic initial conditions (e.g. C$^+$, N, O, Fe$^+$, etc), with pre-existing H$_2$.
HHB10 modeled the chemical evolution as the molecular cloud (the progenitor of the core) forms in a postshock region of atomic gas and suggested that their final chemical abundances
could be a reliable initial condition when the chemical evolution is further calculated in dense cores.
We adopt their results of Model 1 to set the initial abundances for our fiducial model.
The most important initial condition in our work is the oxygen abundances in the atomic gas and the water ice because relative fraction of atomic oxygen in the gas 
will determine the final abundance of newly formed (deuterated) water ice.
This effect will be presented in Section 3.1.

First, we assume that all carbon at the start of our calculations is in the form of gaseous CO with an abundance of $2.8\times10^{-4}$ \citep{lacy, lee03}.
We also assume that half of total oxygen is in CO.
The remaining half of oxygen is found  as water ice or gas phase atomic oxygen.  Based on Model 1 of HHB10, the abundance ratio between atomic oxygen and water ice is about 2.
Therefore, the abundances of atomic oxygen and water ice for our fiducial model are $1.9\times10^{-4}$ and $0.9\times10^{-4}$, respectively.
We assume the initial atomic hydrogen abundance of $1\times 10^{-5}$ because the H density is about 1 \cc\ in molecular clouds \citep{chang07}, and n(H$_2$)$=10^5$ \cc\ in our fiducial model.
(However, the initial H abundance does not affect the final results of our model.)
The initial abundances for our fiducial model are listed in Table~\ref{tab:ini}.

\subsection{Ortho-to-Para Dependent Deuterium Chemistry in the Gas}
The approach adopted here is to assume that the {\em opr}s of key elements (H$_3^+$, H$_2$D$^+$, \DtwoHp , and \Dthreep) are in equilibrium at the gas temperature.  However, the H$_2$ {\em opr} is a free parameter. We then adopt the state to state rate coefficients for each reaction calculated by \citet{hugo09} modified by the spin state fractional amount for both reactants.  For example we take the following reaction:

\begin{equation}
{\rm o\!-\!H_2D^+ + o\!-\!H_2 \rightarrow p\!-\!H_3^+ + HD}.
\end{equation}

\noindent \citet{hugo09} determine a rate of $k_{2} = 4.67 \times 10^{-11} {\rm exp}(0.82/T)$ cm$^{3}$ s$^{-1}$.  
In our network we calculate the equilibrium value of the {\em opr} of \HtwoDp\ for a given temperature while 
the  H$_2$ {\em opr} is provided as an input. Then, we multiply the associated reaction rate by these 
fractions.  In our example above the modified rate, $k_2' = k_2 \times f_{\rm o-H_2D^+} \times f_{\rm o-H_2}$.  
We do not track the {\em opr}s of the products because we {\em a priori} assume the ratios are in equilibrium 
(or fixed for \Htwo). As a result, the rate coefficient for the reaction of ${\rm o\!-\!H_2D^+ + o\!-\!H_2}$ 
becomes  $1.78\times 10^{-17}$ at 10 K with the  H$_2$ {\em opr} of $7\times 10^{-5}$, regardless of the spin type of the product 
(i.e., o--\hthp\ and  p--\hthp\ are not considered separately in the product). 
The rate coefficients ($k_2'$) derived from the state-dependent rates of \citet{hugo09} at 10, 20, and 30 K are listed in Table~\ref{tab:rate}. 

We estimate the {\em opr}s or meta-to-ortho ratios using the energy levels given in the CDMS database \citep{muller05} and the spin statistics given in Table~II of \citet{hugo09}.  The ratio is then calculated by 

\begin{equation}
\frac{\rm ortho}{\rm para} = \frac{2I_o + 1}{2I_p + 1} \times \frac{\sum (2J_o + 1)\; {\rm exp}(-E_o/kT)}{\sum (2J_p + 1)\; {\rm exp}(-E_p/kT)}.
\end{equation}
We list the {\em opr}s and meta-to-ortho ratios at 10, 20, and 30 K at 
Table~\ref{tab:opr}.

In tracking the various ratios, the {\em opr} of H$_2$ remains the most important as it, along with temperature, must be below certain values to allow fractionation to proceed.  
Although it is a free parameter in our model, we also derive the equilibrium value 
of {\em opr}-H$_2$ adopting the usual relation of {\em opr}-H$_2$ = 9 $\times $ exp(-170.5/T) \citep{bourlot91} for below 80 K.  Furthermore,  WFP04 have shown that the \Htwoopr\ does not reach the equilibrium value ({\em opr}-H$_2$ $= 3.5 \times 10^{-7}$) at 10 K. Rather over a wide range of densities, the value reached is instead {\em opr}-H$_2$ $\sim 7 \times 10^{-5}$.   We therefore have fixed the lower bound of the H$_2$ {\em opr} to the latter value.   \citet{flower_dfrac} explored the temperature dependence of the {\em opr} for other key molecular species (H$_3^+$, H$_2$D$^+$, \DtwoHp , \Dthreep ) and find that above $\sim 15$~K the equilibrium ratio is a good approximation.  For gas temperatures below $\sim 15$~K the ratios are generally above the equilibrium value.   Outside of H$_2$, which is the most important player, we do not account for this effect.

\subsection{Grain Surface Chemistry}

In our calculation, two types of grain surface chemistry (formation of \Htwo\, HD, and D$_2$; formation of OH, OD, H$_2$O, and HDO ices) are included. For the surface chemistry, we adopt the method used in \citet{fogel11}, which followed the treatment of \citet{hollenbach09}. For the formation of \Htwo\ and water, an H or O atom must be frozen on the grain surface before another  atom finds it. Therefore, the reaction rate coefficients for the surface chemistry vary with time depending on the abundance ratio between the ice species and grain. 
For example, if the abundance of H atoms on grains (H ice) is greater than the grain abundance, the reaction rate is reduced by the ratio of the grain abundance 
to the abundance of H ice because a gaseous H atom must collide with a grain to initiate the \Htwo\ formation reaction.   
We scale the rate coefficient of each reaction in \citet{fogel11} by the mass ratio between the deuterium atom and the hydrogen atom to treat the formation of HD and deuterated water ice. 
According to Kristensen et al. (2011), the adsorption energies for D$_2$ and HD of 72.0 and 68.7 meV. The ratio of these binding energies is only 1.05, which is very small considering the mass ratio (1.3) between D$_2$ and HD. The mass added to the total mass of a molecule by D is a small fraction, so we assume the same binding energies for both deuterated and normal species, except for D and H, which have the binding energy ratio of 1.2 \citep{perets07}.

\section{Results}
\subsection{Deuterium Fractionation at $T_{\rm gas}$ = 10~K and Low \Htwoopr\ }

Observations of gas-phase molecular depletion, which is dominated by freeze-out onto grain surfaces have suggested that {\em perhaps} all heavy elements are frozen on grain surfaces \citep[see][for a more complete discussion]{bt_araa}.    Under this circumstance the only remaining gas-phase species with a dipole moment that are present are HD, H$_2$D$^{+}$, and D$_2$H$^+$.   Of these HD will not emit appreciably at 10 K, thus observations and several models focused on the question of formation and presence of the two ions in the centers of dense cores (see WFP04).   Below we will compare our results to WFP04 in order to ascertain the similarities/differences between WFP04 and our (similar) chemical model and to isolate the reason of the similarities/differences.

Figure 1a shows the equilibrium abundances of  the key ions as functions of density at the same physical condition as the standard model of WFP04 ($a_g = 0.1$ $\mu$m, T$=$10 K, $\zeta = 3 \times 10^{-17}$ s$^{-1}$) with the complete depletion of heavy elements. 
The level of ionization is comparable between the two models, but we see major differences in the most abundant D-bearing ion.   In our models \Dthreep\ is the most abundant ion for all densities, while WFP04 find a changing distribution switching from H$^+$ (low density) to D$_3^{+}$ (high density) as seen in Figure~2 of WFP04. Identical differences are found in comparison to the work of \citet{flower_dfrac} by  \citet{sipila10}.   They show that in adopting state-to-state coefficients of \citet{hugo09} there is a $\sim$factor of 5  difference in key fractionation reactions
\citep[see][]{sipila10}. 
Our results are not entirely identical with \citet{sipila10}. For example, at \nhtwo\ = $10^5$ \cc, H$_3^+$ is the main ion in our network compared to H$^+$ in theirs.  This is likely due to the fact that we treat our formation of H$_2$ differently by correcting for the possibility that there may be less than one hydrogen per grain \citep{fogel11}. Indeed, H$^+$ becomes the main ion if we turn off our H$_2$ formation reaction. In addition, \citet{sipila13} showed that H$_3^+$ becomes the main ion if the surface chemistry is explicitly included to their calculations \citep[see Figure 3 of ][]{sipila13}.

In Figure 1b, the effect of the surface chemistry of water ice on the {\em gas-phase} deuterium fractionation is presented.  
In this figure, we plot the abundances at a timescale of $10^7$ years for all densities.
The timescale for the chemistry to reach equilibrium varies with the density, i.e., a longer timescale
for a lower density. However, the timescale in our density range is much shorter than $10^7$ years.
Therefore, our models reach similar conditions as WFP04 at $10^7$ years.
Earlier studies suggested the D-bearing ions as main tracers of dense and cold regions \citep{rhm04, aikawa05}. However, as shown in Figure 1b, the abundances of deuterated ions drop at high densities when the surface chemistry of water ice formation is included.  This is due to a decline in the abundance of HD as fractionation progresses and D atoms are locked in ices. In our calculations for n(H$_2$)$\ge 10^6$ cm$^{-3}$, the depletion of deuterated ions is somewhat exaggerated because our model does not consider other surface reactions that can produce HD \citep{sipila13}. 

Figure~2a illustrates the effect of HD depletion, which has been also characterized in \citet{sipila13}.
Deuterium atoms, which are produced in the gas by the dissociative recombination of the deuterated ions, become frozen onto grain surfaces to form deuterated water ice, resulting in the reduction of the HD abundance in the gas.  Therefore, the formation of water ice on grain surfaces significantly affects the {\em gas-phase} deuterium fractionation and changes the main charge carriers. The sharp increase of gas H and D abundances occurs around $10^5$ years because the H-H$_2$
chemistry reaches initial equilibrium at the time. 
Before equilibrium, at low temperatures (T $< 20$ K), the atomic hydrogen is quickly frozen on grain surfaces to form water ice as soon as it is dissociated from other species.

Figure 2b shows the evolution in the abundance of (deuterated) water ice and the D/H ratio in water ice for identical conditions as adopted for the results show in Figure 2a.
Both \HtwoO\ and HDO abundances increase with time, with a steeper gradient in HDO resulting in the increase of the D/H ratio with time.  The steeper gradient leading to HDO ice formation is a direct result of the rise in the D/H ratio in the atomic pool (Figure~2a).
The calculated maximum D/H ratio of water ice is about 0.05, which should be considered as the upper limit, because we only consider the water surface chemistry.    In a more expansive model the deuterium atoms could be incorporated into other deuterated ices \citep{tielens83}; although water ice still dominates.

\citet{sipila13} also used the rates of \citet{hugo09} and incorporated the surface chemistry explicitly into their 
calculation, so it makes a good comparison with our work.  We adopted their initial abundances to calculate 
the chemical evolution in the model with $n(\rm H_2)=10^5$ cm$^{-3}$ and $T_{\rm gas}=10$ K, which are similar to the 
conditions of the outermost shell in their model core. We also used their cosmic 
ray ionization rate ($\zeta=1.3\times10^{-17}$ s$^{-1}$), but our constant {\em opr}s and meta-to-ortho ratios 
at 10 K are adopted for the calculation. 

In the calculation of \citet{sipila13} the {\em oprs} vary with time; however, except for H$_2$, the variation in opr is within an order of magnitude.  This does not produce a notable change in the resulting abundances of \Hthreep\ and its deuterated ions as well as the H$_2$O/HDO ices.
Figure 3 can be compared with Figure 4 (the right panel) of \citet{sipila13} directly.  
The overall trends are very similar although the abundances reach the steady state after $\sim 10^6$ years because only water ice formation is included in our calculation. The biggest difference between this test and our fiducial model is the D/H ratio of water; in this test, it reaches 0.13, about three times higher than the D/H ratio in the fiducial model. This is because all oxygen in \citet{sipila13} is initially atomic and available to eventually be incorporated into HDO and H$_2$O on grain surfaces.

 We tested different initial abundance ratios between water ice and gaseous atomic oxygen at two temperatures, 10 and 20 K. The equilibrium H$_2 ~opr$ at each temperature was adopted. 
 The result is presented in Figure 4. The D/H ratio of water ice at the timescale of $3\times 10^5$ years depends on the relative disposition of water ice and atomic oxygen at the start of our calculation.  For a higher initial abundance of water ice relative to atomic oxygen, we find a lower water D/H ratio on grain surfaces.  This is because if more oxygen is locked initially 
 in water ice, less oxygen is available to form H$_2$O and HDO on grain surfaces at later times, resulting in a reduced water D/H ratio.
 In addition, because of the active backward reactions of deuterium fractionation at the higher temperature of 20 K, the overall D/H ratio of water ice is lower than that at 10 K.
 
Recently, \citet{taquet13a} developed a full deuterium chemical network, adopting explicit surface 
chemical reactions for the formation of various molecular ices as well as water ice, to show that
other molecular ices such as CH$_3$OH and H$_2$CO  can be more deuterated than the water ice. 
However, the water ice is the most abundant ice on grain surfaces, with an abundance 
higher than other ices by orders of magnitude in most conditions as shown in \citet{taquet13a}. 
Therefore, the biggest reservoir of deuterium in grain 
surfaces is water ice, which justifies our focus on water surface chemistry. Figure~5 shows the D/H ratio
of water ice versus the density, which has been calculated with the same parameters used in Figure~10 
of \citet{taquet13a}, i.e., the H$_2$ {\em opr} of $3\times 10^{-6}$ and $A_v=10$ mag. For the calculation, 
we adopted our fiducial initial abundances. The results of our calculations are very consistent with those 
by \citet{taquet13a}, and the D/H ratio of water vapor observed in the hot corino of IRAS 16293
 \citep{coutens12} is well explained by our model with this H$_2$ {\em opr}. 
 \citet{dislaire12} determined that the  H$_2$ {\em opr} is $1\times 10^{-3}$ in the dense core of IRAS16293.  With this ratio we can also explain the observed D/H ratio water as presented dotted lines in Figure 5; if the temperature is 20 K, the density of IRAS 16293 must be higher than $10^5$ cm$^{-3}$.

\subsection{Deuterium Fractionation For Variable Gas Temperature, \Htwoopr\, and Density }

The D/H ratio of water ice is a function of the efficiency of deuterium fractionation in the gas
which in turn depends on the \Htwoopr\ {\em and} the gas temperature.   In the dense ISM there exist a range of environments where the ambient dense star-forming material has gas temperatures of $\sim 10$~K (e.g. Taurus), but other regions have higher values (e.g. Orion).
Figure~6 shows the two-dimensional dependence for the D/H ratio of water ice and HD from T$_{gas}  = 10 - 30$~K and for \Htwoopr\ from $\sim 10^{-5}$ to the high temperature ratio of 3.
In Figure 6  the fractionation ratio is calculated at the density of $10^5$ \cc\ with our fiducial initial abundances at the timescale of $3\times 10^5$ yrs. 
In this calculation, the \Htwoopr\ is set to be a constant as a free parameter, independent of temperature.   
The white dotted line in Figure 6 indicates the equilibrium \Htwoopr\ calculated by the relation, $ 9 \times exp(-170.5/T)$. 
The equilibrium value becomes lower than $7\times 10^{-5}$ at T$\lesssim$15 K. In the case, we plot $7\times 10^{-5}$. The D/H ratios along the white lines are the values when \Htwoopr\ reaches the equilibrium at given temperatures.

Clearly in Figure~6 there is a strong dependence on the D/H ratio for water ice on both parameters.  However, the D/H ratio is insensitive to the \Htwoopr\ when the \Htwoopr\ is smaller than $\sim 10^{-3} - 10^{-2} $.      
\citet{flower06} showed that the \Htwoopr\ decreases with collapse, while we simply explore the effects of fractionation for constant \Htwoopr .
However, the D/H ratio of water ice at a given temperature is not sensitive to the \Htwoopr\ as long as it is smaller than $10^{-2}$, which is true at all times in the Flower et al. collapse model.  Moreover, the formation of the molecular cloud itself will precede the formation of the dense core.   Thus there is potentially $\sim 10^6 - 10^7$ yrs for the gas to evolve to a lower \Htwoopr\ ratio (and setting the stage for subsequent D enrichments) prior to the condensation and collapse of a dense molecular core \citep[e.g.][]{bergin_cform, clark12}.   

Over this range the ratio of HD/H$_2$ shows an inverse dependence (Figure 6, bottom).
That is, the D/H ratio of molecular hydrogen increases with the gas temperature and \Htwoopr , trending towards the cosmic value.  
When  \Htwoopr\ is $< 10^{-2}$ and T$_{gas} < 20$~K, the abundance of HD exhibits depletions.
The lowest HD/${\rm H_2}$ ratio from this calculation at $\nhtwo=10^5$ \cc\ is $\sim 7.8\times 10^{-6}$, which is lower than the comic ratio ($\sim 3\times 10^{-5}$) by a factor of about 4.  
The fraction is even lower ($\sim 2\times 10^{-7}$) by more than two orders of magnitude compared to the cosmic ratio at $\nhtwo=3\times 10^5$ \cc.

Figure 7 presents the abundance of \Hthreep\ and its deuterated isotopologues across the identical two dimensional grid. 
\Hthreep\ is always the most dominant ion. For a temperature lower than $\sim 20$ K, the freeze-out of deuterium as ices produces low abundance ratios of deuterated ions relative to \Hthreep .  Furthermore the ratios drop sharply at higher temperatures due to the activation of the backward reactions.
In our fiducial model, the grain mantle is coated with water ice, so the binding energies of molecules are greater compared to those to the bare silicate grain.
As a result, even at 30 K, most CO, which is a primary destroyer of H$_3^+$ and H$_2$D$^+$ via the reactions of ${\rm H_3^+ + CO \rightarrow HCO^+ + H_2}$ 
and ${\rm H_2D^+ + CO \rightarrow DCO^+ + H_2}$, is still frozen on grain surfaces, leaving H$_3^+$ abundant.
However, in the model with ${\rm T_{gas}=25}$ K with no initial water ice (i.e., with binding energies to the bare silicate grain), CO evaporates and the abundance of H$_3^+$ is lowered to $10^{-9}$.

Figure 8 shows the dependence of the D/H ratio of water ice and the H$_2$ gas on the temperature and density when the equilibrium \Htwoopr, which is marked as  the white dashed lines in Figure 6 and 7, is assumed at each given temperatures. In the densities greater than $\sim 3\times 10^5$ \cc, the D/H ratio is only dependent on the gas temperature. The highest D/H ratio of water ice is almost 0.1 for densities greater than $3\times 10^5$ \cc\ and T$\le 15$ K. As noted above, this value must be considered as a maximum ratio since we considered only the water ice formation and did not include other potential deuterium carriers (e.g., H$_2$CO, CH$_3$OH, and NH$_3$).  However,  water ice is still the most abundant ice and the major reservoir of deuterium in the icy grain mantle. 

Figure 9 presents the abundance distribution of H$_3^+$ and its deuterated species in the  same domain of temperature and density as used in Figure 8.
H$_3^+$ is not sensitive to temperature as seen in Figure 9 and predominantly exhibits a density dependence due to the increase recombination rates.   The abundance of the deuterated daughter products of H$_3^+$ are  sensitive to density at temperatures lower than 20 K. At gas temperatures greater than 20 K, their abundances drop sharply and are almost constant. This is due to the activation of the backward reactions which slows down deuterium fractionation for T$> 20$ K. 
The effective formation of the deuterated water ice on grain surfaces at higher densities results in the significant depletion of HD in the gas (Figure 8), which in turn results 
in the heavy depletion of deuterated ions at the cold and densest regions (Figure 9). 
Figure 10 shows the abundance ratio between H$_2$D$^+$ and H$_3^+$, which presents a similar trend to the explained above; the abundance ratio changes very sharply around $3\times 10^5$ \cc at T$\le 20$ K, but at T$>20$ K, the abundance ratio drops with temperature and is insensitive to density.

\section{Discussion and Implications}

It is clear from numerous investigations discussed above that the deuterium chemistry has an innate dependence on the ortho to para ratio of H$_2$, the gas temperature, and the gas density.   In this paper we have attempted to provide an initial
look at the interdependence of these key parameters in terms of the overall initiation of deuterium chemistry in the gas, but also the deuterium implantation into ices.   We note that we are not alone in exploring this parameter space and we also refer the reader to \citet{taquet13a} and \citet{sipila13}.  Below we attempt to use observations to provide additional perspective.

\subsection{Water ice}

A wide range of D/H ratios of  water vapor  have been reported in the hot gas near protostars from $< 6\times 10^{-4}$ to $>$ 0.01 \citep{jacq90, gmw96, helmich96, stark04, parise05, bergin10a, jj10, liu11, coutens12, neill13a, persson13}.   These values are believed to trace the deuterium enrichment of evaporated water ice which formed during earlier colder phases.
In addition, \citet{coutens12} argued that the water should form before the gravitational collapse of a protostar based on their calculation of nearly constant water D/H ratios ($\sim 0.01$) from the hot core to the low density outer envelope of IRAS 16293-2422. 
Thus, one interpretation of the D/H ratios is 
that these variations reflect differences in the initial conditions, in particular the \Htwoopr, gas temperature, and density, (and also cosmic ray ionization rate), under which the ices form.     Thus to account for this range one possibility is that the \Htwoopr\ could vary from source-to-source or even along the line of sight, while assuming that the gas generally remains cold (T $< 15$~K).  It is generally believed that H$_2$ forms on grains with an $opr$ of 3, which is then gradually thermalized towards the ratio at the local gas temperature via reactions with H$^+$ and H$_3^+$.    The interconversion of ortho and para-H$_2$ is slow \citep[requiring timescales $>$ 3 $\times$ 10$^6$ yrs for steady state;][]{flower06}.  However, \citet{pagani09, pagani11} demonstrate that the conversion is dependent on the local gas density with order of magnitude variations in the \Htwoopr.   Thus differences in the density evolution during the core formation phase can result in variations in the $opr$ of H$_2$.   Clearly this is complicated by the fact that the gas temperature is also not constant and can vary from source-to-source.

If we focus on the extremes the high enrichment levels $>$ 0.01, such as towards the low-mass protostar IRAS16293 \citep{coutens12},  require both low \Htwoopr\ $< 0.01$ and low gas temperatures ($<$ 20~K) in our models.    Moreover the gas would need to stay within these conditions for $\sim 10^5$ yrs., which appears plausible given current timescales for core formation \citep[see discussion in][]{pagani11}.    
Low ratios, such as found towards the low-mass protostar NGC1333 IRAS 4B \citep{jj10}, cannot be achieved unless the gas temperature during much of the evolution is $>$ 15 K while also requiring an \Htwoopr\ $>$ 0.1.       Thus our models, with a different set of parameters, can match the observed range.  However, both of these sources are in low-mass star-forming regions where the ambient gas is cold with characteristic temperatures of $\sim 10-15$~K \citep{bt_araa}.  Moreover it would require an \Htwoopr\ that varies by over 2 orders of magnitude, despite the  roughly similar conditions of ambient material.    In this light perhaps not all measured D/H ratios are intrinsic (i.e. set at birth) and some mechanisms could be operative to alter the ratio in  the very hot gas near protostars \citep[see discussion in][]{neill13a}.  However, as suggested by \citet{jj10} higher resolution observations can help in providing a more meaningful comparison between the various measurements.

We note that attempts to detect solid HDO in YSOs have been also made \citep{dartois03, parise03}  setting an upper limit of the D/H ratio of $\lesssim 0.02$ in water ice.
Considering the parameters associated with the surface chemistry such as binding energies, thse limes are consistent with our predictions.  The highest equilibrium D/H ratio of water ice in our fiducial model is  $\sim 0.05$ at 10 K and $10^5$ \cc. 
This value is a firm upper limit because we have a limited gas-phase network that explores water ice formation.  Thus we do not  consider hydrogenation of other ices such as 
H$_2$CO, CH$_3$OH, and NH$_3$.

\subsection{Ions in the cold gas}

As summarized in the review of dense cold cores by \citet{bt_araa} 
dense pre-stellar molecular cores often exhibit a high degree of central concentration. 
Furthermore most of their mass resides at a low temperature of $\sim 10$~K.   Under these conditions chemical models have previously predicted that \HtwoDp , \DtwoHp , and \Dthreep\ would be excellent tracers of the dense core center where many other molecules will be frozen onto grain surfaces \citep{rhm04, aikawa05}.    The dense protoplanetary disk midplane presents similar cold dense environment and \citet{cc05} predict that the deuterated forms of \Hthreep\ will be the primary charge carriers.  However, these studies considered only the depletion of molecules without dealing with actual surface chemistry. 
Because HD is not frozen on grain surfaces, 
in the condition of the high degree of molecular depletion, the deuterated ions would be the main charge carriers in the gas. 

However, D atoms, produced by the dissociative recombination of deuterated ions, are also frozen on grains surfaces to eventually be incorporated into molecular ices. 
Although our study considered only water ice formation, it can catch the representative effect of the depletion of deuterium, leaving \Hthreep\ as the primary species
in the gas.
If our results are correct, therefore, the \HtwoDp\ abundance of $10^{-10} \sim 10^{-9}$ derived by \citet{caselli03} might correspond to the region with the density of 1 to $3\times 10^5$ \cc\ rather than $\sim 10^6$ \cc, where almost all deuterium is converted to the deuterated water ice as seen in Figure~1b (X(\HtwoDp) $\le 10^{-12}$).  Thus, molecular ions like \HtwoDp ,\DtwoHp , \Dthreep\ would not trace core centers or the dense midplane of protoplanetary disks, but rather dense intermediate layers that are closer to the core or disk surface where HD can exist with higher abundance.
According to \citet{caselli08}, which surveyed dense starless/protostellar cloud cores in the ${\rm o\!-\!H_2D^+}$ ($1_{1,0}-1_{1,1}$) line, the column density of 
${\rm o\!-\!H_2D^+}$ is not constrained solely by one parameter such as CO depletion factor or density. As presented in this study, the abundances of ions are 
functions of density, temperature, \Htwoopr , and the gas-grain interaction.

Nevertheless, HD is  destroyed significantly only at n(H$_2$)$\ge 3\times 10^5$ \cc\ and
T$< 20$ K (Figure 8). For instance, the abundance of \HtwoDp\ is about $2\times 10^{-9}$ at $\nhtwo=10^6$ \cc\ and T$=20$ K (Figure 9). Therefore, deuterium bearing ions will still trace the densest gas at T$\ge 20$ K.

\acknowledgments

We thank the anonymous referee, whose comments led to improvements in the paper.
This research was supported by the Basic Science Research Program
through the National Research Foundation of Korea (NRF) funded by the Ministry of
Education of the Korean government (grant No. NRF-2012R1A1A2044689).   
This work was also supported by the National Science Foundation under Grant \#1008800 and by NASA under grant NN08AH23G from the Astrophysics Theory and Fundamental Physics and Origins of Solar System programs.

\begin{deluxetable}{lr}
\tablecaption{Initial Abundances
\label{tab:ini}}
\tablewidth{0pt}
\tablehead{
\colhead{Species} &
\colhead{Abundance (Relative to \Htwo)}
}
\startdata
H & $1.00\times10^{-5}$\\
He & 0.18\\
He$^+$ & $1.33\times 10^{-10}$\\
CO & $2.80\times 10^{-4}$\\
HD & $2.80\times 10^{-5}$\\
H$_2$O (ice) & $9.00\times 10^{-5}$\\
O &$1.90\times 10^{-4}$\\
N & $4.28\times 10^{-5}$\\
SI$^+$ & $3.40\times 10^{-9}$\\
Mg$^+$& $2.20\times 10^{-9}$\\
S$^+$& $5.60\times 10^{-7}$\\
Fe$^+$& $3.40\times 10^{-9}$
\enddata
\end{deluxetable}

\begin{deluxetable}{lccc}
\tablecolumns{4}
\tablecaption{Rate Coefficients
\label{tab:rate}}
\tablewidth{0pc}
\tablehead{
\colhead{}    &  \multicolumn{3}{c}{Rate Coefficients} \\
\cline{2-4} \\
\colhead{Reactions} & \colhead{10 K}   & \colhead{20 K}    & \colhead{30 K} }
\startdata
 p-\hdtp\ +    p-\dt\ \ra\    \dthp\ +        HD & 2.61E-15 & 2.17E-12 & 1.63E-11 \\
 p-\hdtp\ +    o-\dt\ \ra\    \dthp\ +        HD & 1.18E-11 & 1.32E-10 & 2.37E-10 \\
 o-\hdtp\ +    p-\dt\ \ra\    \dthp\ +        HD & 2.65E-13 & 1.73E-11 & 6.01E-11 \\
 o-\hdtp\ +    o-\dt\ \ra\    \dthp\ +        HD & 1.13E-09 & 9.97E-10 & 8.25E-10 \\
 m-\dthp\ +        HD \ra\    \hdtp\ +      \dt\ & 5.32E-17 & 9.95E-14 & 1.07E-12 \\
 o-\dthp\ +        HD \ra\    \hdtp\ +      \dt\ & 7.46E-17 & 1.35E-13 & 1.69E-12 \\
 o-\dthp\ +    p-\hh\ \ra\    \htdp\ +      \dt\ & 2.23E-25 & 5.52E-18 & 1.48E-15 \\
 o-\dthp\ +    p-\hh\ \ra\    \hdtp\ +        HD & 8.77E-20 & 1.14E-14 & 5.61E-13 \\
 o-\dthp\ +    o-\hh\ \ra\    \htdp\ +      \dt\ & 7.67E-26 & 7.70E-19 & 9.03E-16 \\
 o-\dthp\ +    o-\hh\ \ra\    \hdtp\ +        HD & 2.23E-16 & 1.27E-13 & 5.86E-12 \\
 p-\htdp\ +    p-\dt\ \ra\    \hdtp\ +        HD & 3.28E-13 & 2.09E-11 & 5.94E-11 \\
 p-\htdp\ +    p-\dt\ \ra\    \dthp\ +      \hh\ & 4.81E-14 & 2.87E-12 & 8.02E-12 \\
 p-\htdp\ +    o-\dt\ \ra\    \hdtp\ +        HD & 9.75E-10 & 8.43E-10 & 5.74E-10 \\
 p-\htdp\ +    o-\dt\ \ra\    \dthp\ +      \hh\ & 4.05E-10 & 3.37E-10 & 2.26E-10 \\
 o-\htdp\ +    p-\dt\ \ra\    \hdtp\ +        HD & 6.19E-16 & 3.18E-12 & 3.55E-11 \\
 o-\htdp\ +    p-\dt\ \ra\    \dthp\ +      \hh\ & 7.36E-17 & 3.82E-13 & 4.28E-12 \\
 o-\htdp\ +    o-\dt\ \ra\    \hdtp\ +        HD & 2.06E-12 & 1.46E-10 & 3.93E-10 \\
 o-\htdp\ +    o-\dt\ \ra\    \dthp\ +      \hh\ & 5.14E-13 & 3.37E-11 & 8.78E-11 \\
 p-\hdtp\ +        HD \ra\    \htdp\ +      \dt\ & 3.97E-16 & 8.49E-14 & 5.11E-13 \\
 p-\hdtp\ +        HD \ra\    \dthp\ +      \hh\ & 5.36E-12 & 6.27E-11 & 1.20E-10 \\
 o-\hdtp\ +        HD \ra\    \htdp\ +      \dt\ & 4.28E-16 & 8.84E-14 & 5.43E-13 \\
 o-\hdtp\ +        HD \ra\    \dthp\ +      \hh\ & 7.47E-10 & 6.58E-10 & 5.75E-10 \\
 m-\dthp\ +    p-\hh\ \ra\    \htdp\ +      \dt\ & 1.68E-25 & 4.27E-18 & 1.03E-15 \\
 m-\dthp\ +    p-\hh\ \ra\    \hdtp\ +        HD & 3.90E-20 & 4.96E-15 & 2.04E-13 \\
 m-\dthp\ +    o-\hh\ \ra\    \htdp\ +      \dt\ & 4.92E-26 & 5.30E-19 & 5.56E-16 \\
 m-\dthp\ +    o-\hh\ \ra\    \hdtp\ +        HD & 1.48E-16 & 8.25E-14 & 3.23E-12 \\
 o-\hdtp\ +    o-\hh\ \ra\    \hthp\ +      \dt\ & 2.56E-23 & 4.05E-18 & 1.21E-15 \\
 o-\hdtp\ +    o-\hh\ \ra\    \htdp\ +        HD & 1.19E-15 & 6.50E-14 & 1.62E-12 \\
 o-\hdtp\ +    p-\hh\ \ra\    \hthp\ +      \dt\ & 1.19E-25 & 1.20E-18 & 2.31E-16 \\
 o-\hdtp\ +    p-\hh\ \ra\    \htdp\ +        HD & 1.29E-18 & 1.91E-14 & 4.35E-13 \\
 p-\hdtp\ +    o-\hh\ \ra\    \hthp\ +      \dt\ & 1.29E-27 & 4.62E-20 & 7.28E-17 \\
 p-\hdtp\ +    o-\hh\ \ra\    \htdp\ +        HD & 4.62E-17 & 2.24E-14 & 1.11E-12 \\
 p-\hdtp\ +    p-\hh\ \ra\    \hthp\ +      \dt\ & 7.83E-28 & 4.65E-19 & 3.21E-16 \\
 p-\hdtp\ +    p-\hh\ \ra\    \htdp\ +        HD & 3.71E-18 & 4.12E-14 & 7.87E-13 \\
 o-\htdp\ +        HD \ra\    \hthp\ +      \dt\ & 2.25E-17 & 1.28E-13 & 1.71E-12 \\
 o-\htdp\ +        HD \ra\    \hdtp\ +      \hh\ & 2.03E-12 & 1.41E-10 & 3.97E-10 \\
 p-\htdp\ +        HD \ra\    \hthp\ +      \dt\ & 3.91E-18 & 5.33E-15 & 4.59E-14 \\
 p-\htdp\ +        HD \ra\    \hdtp\ +      \hh\ & 1.35E-09 & 1.20E-09 & 8.78E-10 \\
 p-\hthp\ +        HD \ra\    \htdp\ +      \hh\ & 1.40E-09 & 1.15E-09 & 1.00E-09 \\
 o-\hthp\ +        HD \ra\    \htdp\ +      \hh\ & 1.04E-10 & 4.17E-10 & 5.96E-10 \\
 p-\htdp\ +    p-\hh\ \ra\    \hthp\ +        HD & 3.76E-20 & 2.64E-15 & 7.98E-14 \\
 p-\htdp\ +    o-\hh\ \ra\    \hthp\ +        HD & 7.96E-17 & 1.39E-13 & 7.78E-12 \\
 o-\htdp\ +    p-\hh\ \ra\    \hthp\ +        HD & 2.04E-19 & 1.74E-14 & 5.36E-13 \\
 o-\htdp\ +    o-\hh\ \ra\    \hthp\ +        HD & 1.78E-17 & 3.96E-14 & 2.01E-12 \\
 p-\hthp\ +    p-\dt\ \ra\    \htdp\ +        HD & 2.25E-12 & 1.25E-10 & 4.08E-10 \\
 p-\hthp\ +    p-\dt\ \ra\    \hdtp\ +      \hh\ & 1.77E-13 & 1.01E-11 & 3.32E-11 \\
 p-\hthp\ +    o-\dt\ \ra\    \htdp\ +        HD & 4.91E-10 & 3.80E-10 & 3.00E-10 \\
 p-\hthp\ +    o-\dt\ \ra\    \hdtp\ +      \hh\ & 9.65E-10 & 7.26E-10 & 5.68E-10 \\
 o-\hthp\ +    p-\dt\ \ra\    \htdp\ +        HD & 1.54E-14 & 4.45E-12 & 2.50E-11 \\
 o-\hthp\ +    p-\dt\ \ra\    \hdtp\ +      \hh\ & 1.46E-14 & 4.20E-12 & 2.36E-11 \\
 o-\hthp\ +    o-\dt\ \ra\    \htdp\ +        HD & 3.05E-11 & 1.35E-10 & 1.89E-10 \\
 o-\hthp\ +    o-\dt\ \ra\    \hdtp\ +      \hh\ & 7.85E-11 & 2.94E-10 & 3.89E-10 \\
\enddata
\end{deluxetable}

 \begin{deluxetable}{rccc}
\tablecaption{Spin Isotopomer Ratios
\label{tab:opr}}
\tablewidth{0pt}
\tablehead{
\colhead{Species} &
\colhead{10 K} &
\colhead{20 K} &
\colhead{30 K} 
}
\startdata
H$_2$ o/p& 7.00 $\times 10^{-5}$ & 1.79$\times 10^{-3}$ & 3.06$\times 10^{-2}$ \\
D$_2$ o/p & 3.62$\times 10^3$ & 49.1 & 11.7 \\
H$_3^+$ o/p & 7.53$\times 10^{-2}$ & 0.388 & 0.663\\
H$_2$D$^+$ o/p & 1.85$\times 10^{-3}$ & 0.152 & 0.599\\
D$_2$H$^+$ o/p & 97.7  & 7.47 & 3.43 \\
D$_3^+$ o/m& 2.30 $\times 10^{-2}$ & 0.244 & 0.581
\enddata
\end{deluxetable}

\begin{figure}
\figurenum{1}
\epsscale{0.6}
\plotone{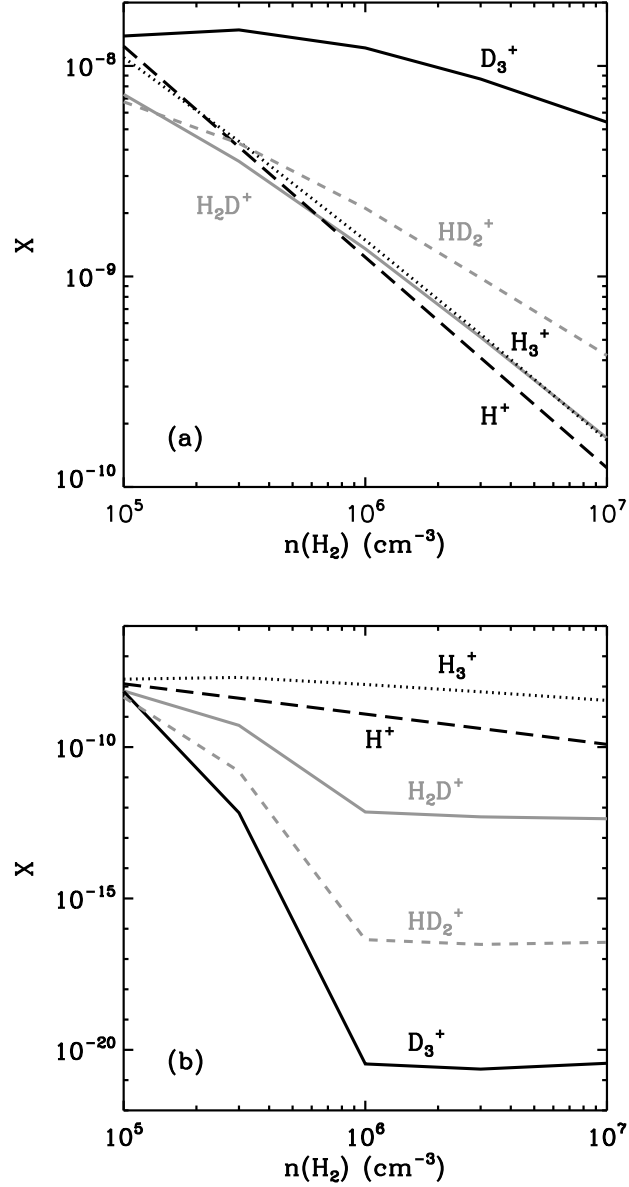}
\figcaption{The effect of water ice formation on the gas-phase deuterium fractionation; (a) the abundances of key ions as functions of density when all heavy elements are depleted and no further water ice formation occurs, and (b) the same as (a) but the surface chemistry of water ice formation is included. The timescale for the plot is $10^7$ years. The extremely low abundances at $n(\rm H_2) \ge 10^6$ \cc\ are due to the complete depletion of HD (our model does not include the reactions to produce HD on grain surfaces.)} 
\end{figure}

\begin{figure}
\figurenum{2}
\epsscale{0.6}
\plotone{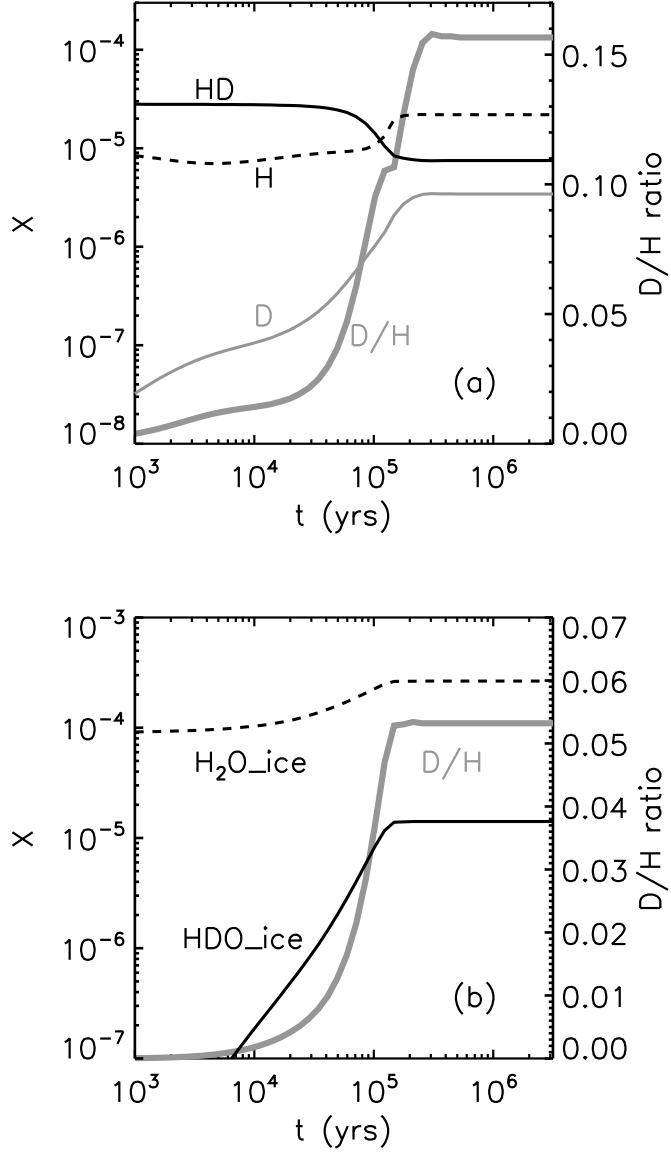}
\figcaption{The evolution of (a) the abundances of HD, H, and D as well as the D/H ratio of the atomic hydrogen gas and (b) the abundances of water ice and deuterated water ice as well as the D/H ratio of water ice in the physical conditions of $\rm T_{gas}=10$ K and $\nhtwo=10^5$ \cc\ with the fiducial initial condition.
}
\end{figure}

\begin{figure}
\figurenum{3}
\epsscale{0.6}
\plotone{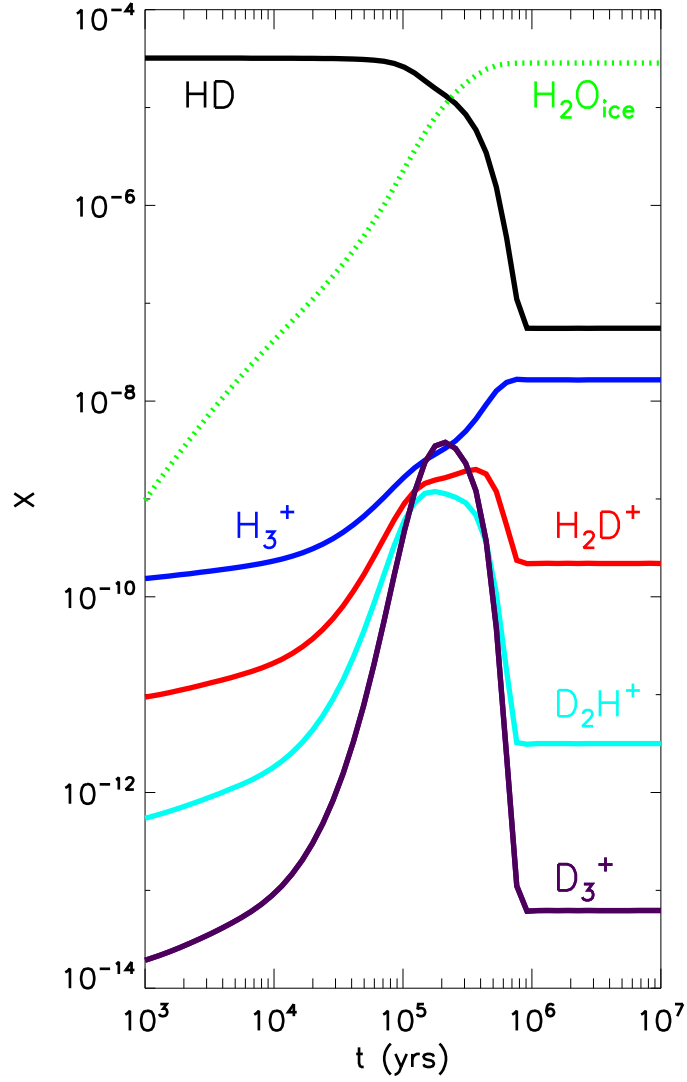}
\figcaption{The abundances (relative to \Htwo) of selected species as a function of time for comparison with \citet{sipila13}. The physical conditions of the model are $T_{\rm gas}=10$ K and $\nhtwo=10^5$ \cc, and the same initial abundances and cosmic ray ionization rate as used in \citet{sipila13} were adopted .  
}
\end{figure}

\begin{figure}
\figurenum{4}
\epsscale{0.6}
\plotone{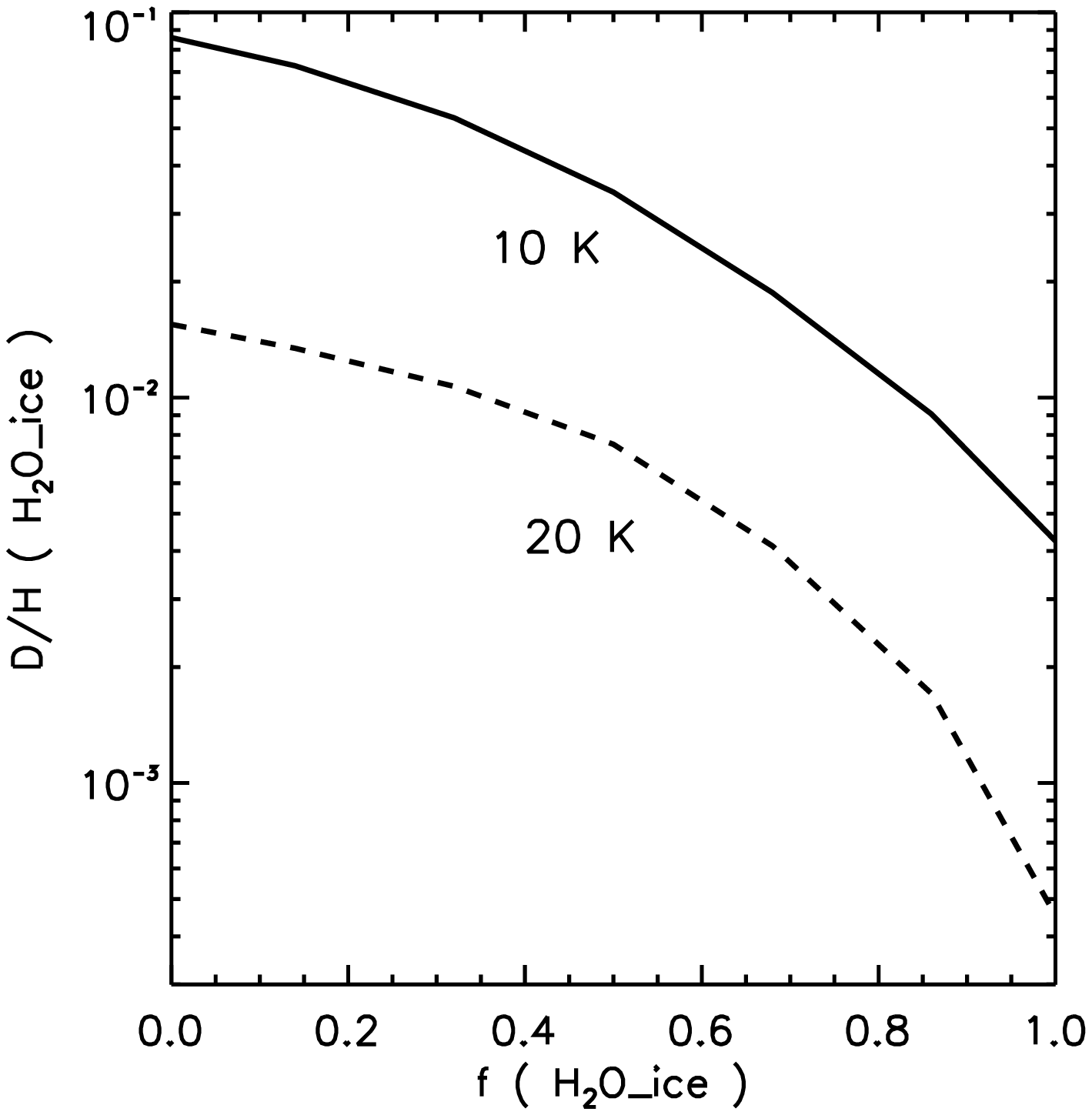}
\figcaption{The effect of the initial water ice abundance. f(H$_2$O\_ice) is the initial ratio between the water ice abundance and the total oxygen abundance excluding oxygen in CO 
(f(H$_2$O\_ice)$=$X(H$_2$O\_ice)/[X(H$_2$O\_ice)$+$X(O\_atom)]). In our fiducial model, f(H$_2$O\_ice)$=0.3$. The timescale for the plot is $3\times 10^5$ yrs.
The sold and dotted lines present the D/H ratio of water ice  when the gas temperature is 10 K and 20 K, respectively. The density is the same as $10^5$ \cc\ for all models .
If more oxygen is locked initially in water ice, the final D/H ratio of water ice becomes lower. Only for the model of f(H$_2$O\_ice)$=0$, i.e., with no initial water ice,
we adopt the binding energies of molecules to the bare silicate grain, which are lower than those to the water ice coated grain by a factor of 1.47. 
}
\end{figure}

\begin{figure}
\figurenum{5}
\epsscale{1.0}
\plotone{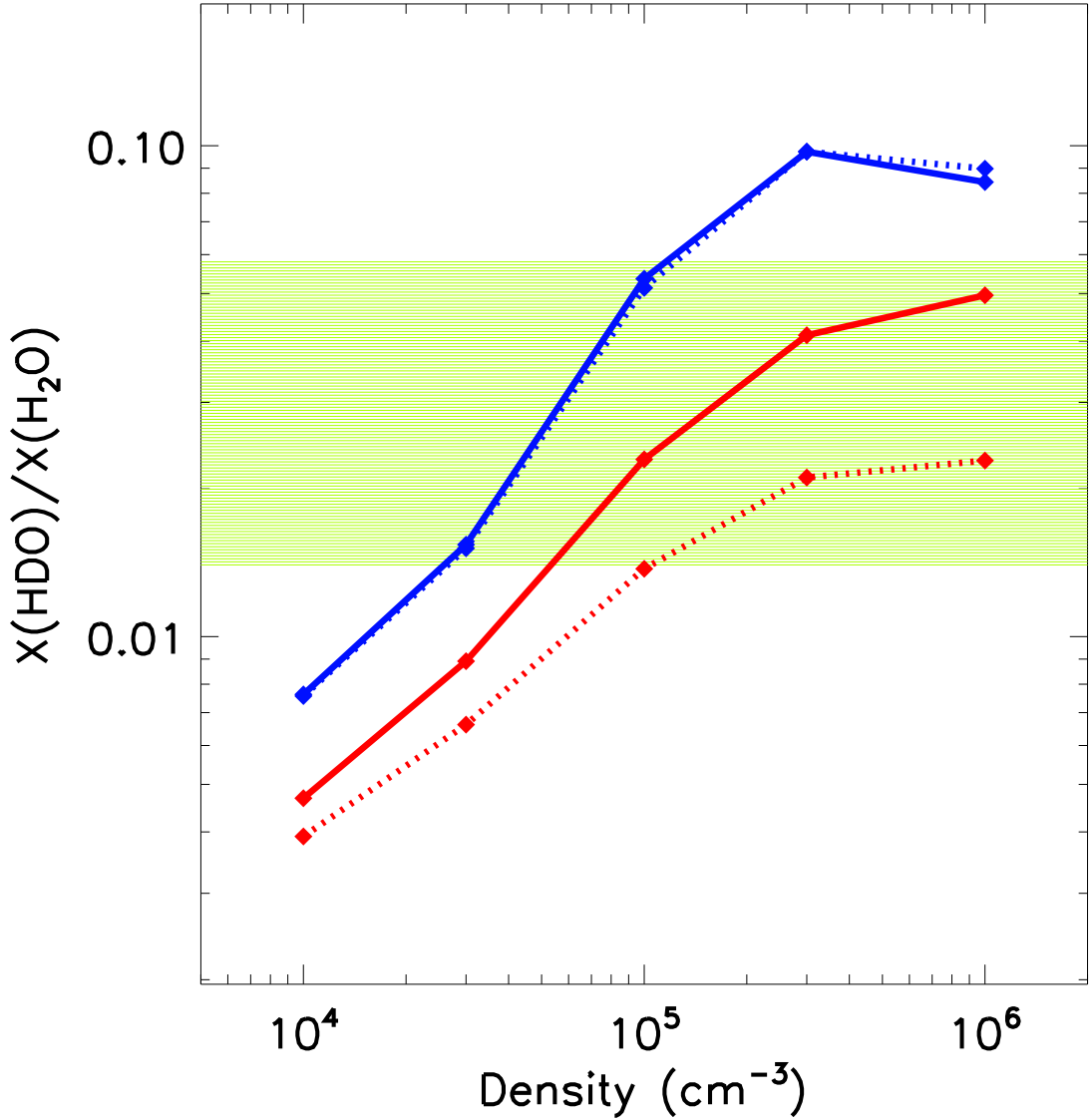}
\figcaption{The D/H ratio of water ice versus the density at $3\times 10^5$ yr. 
The same parameters (H$_2$ {\em opr}$=3\times 10^{-6}$ and $A_v=10$ mag) as used in Figure 10 of
\citet{taquet13a} have been adopted (solid lines). The blue and red lines indicate the results at 10 K and 20 K,
respectively. The green box refers to the D/H ratio of water observed in the hot corino of IRAS 16293 \citep{coutens12}. The dotted lines are for H$_2$ {\em opr}$=1\times 10^{-3}$.} 
\end{figure}

\begin{figure}
\figurenum{6}
\epsscale{0.7}
\plotone{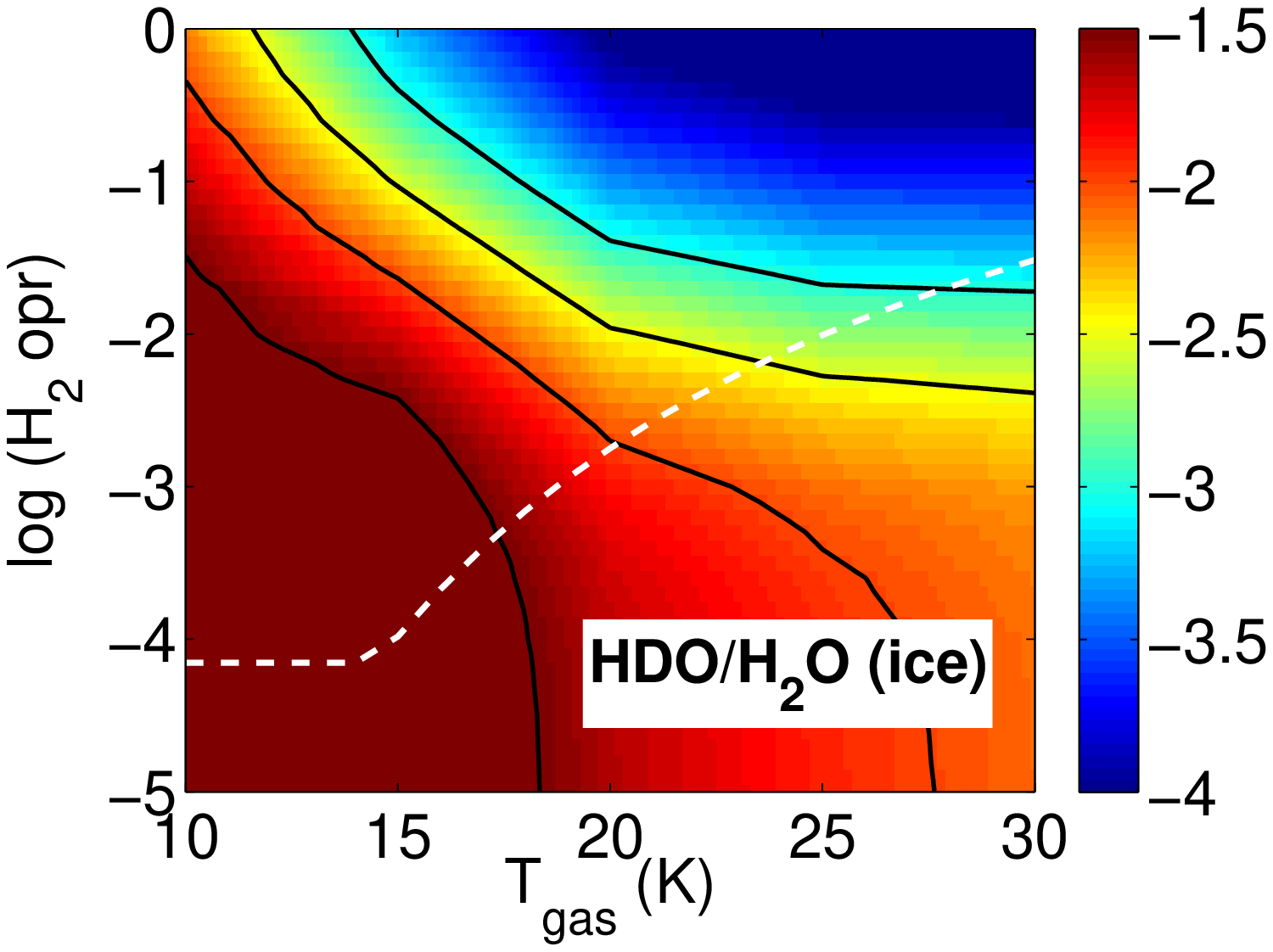}
\plotone{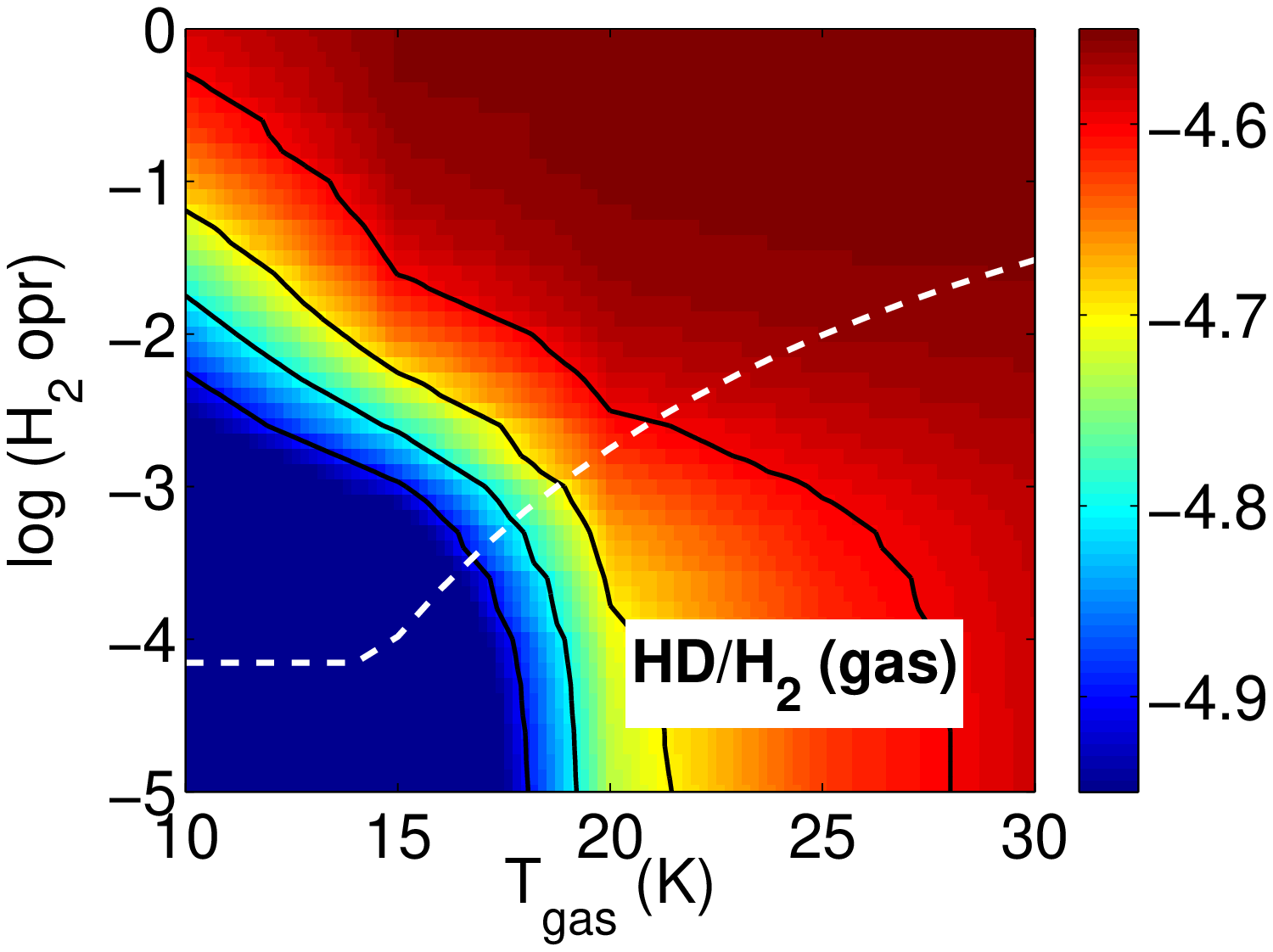}
\figcaption{The D/H ratio ($log({\rm D/H})$) of water ice (top) and molecular hydrogen gas (bottom) in the two dimensional grid of the gas temperature and \Htwoopr\ in $\nhtwo=10^5$ \cc\ at the timescale of $3\times 10^5$ yrs. The dust temperature is the same as the gas temperature in our models. The fiducial initial abundances were adopted for the calculation. The contours represent $log({\rm D/H})$ of -1.5, -2, -2.5, and -3 for water ice and -4.6, -4.7, -4.8 and -4.9 for molecular hydrogen gas from the red to blue colors. The white dashed line indicates the equilibrium \Htwoopr\ above $\sim$15 K and $7\times 10^{-5}$ below 15 K. 
}
\end{figure}

\begin{figure}
\figurenum{7}
\epsscale{1.0}
\plottwo{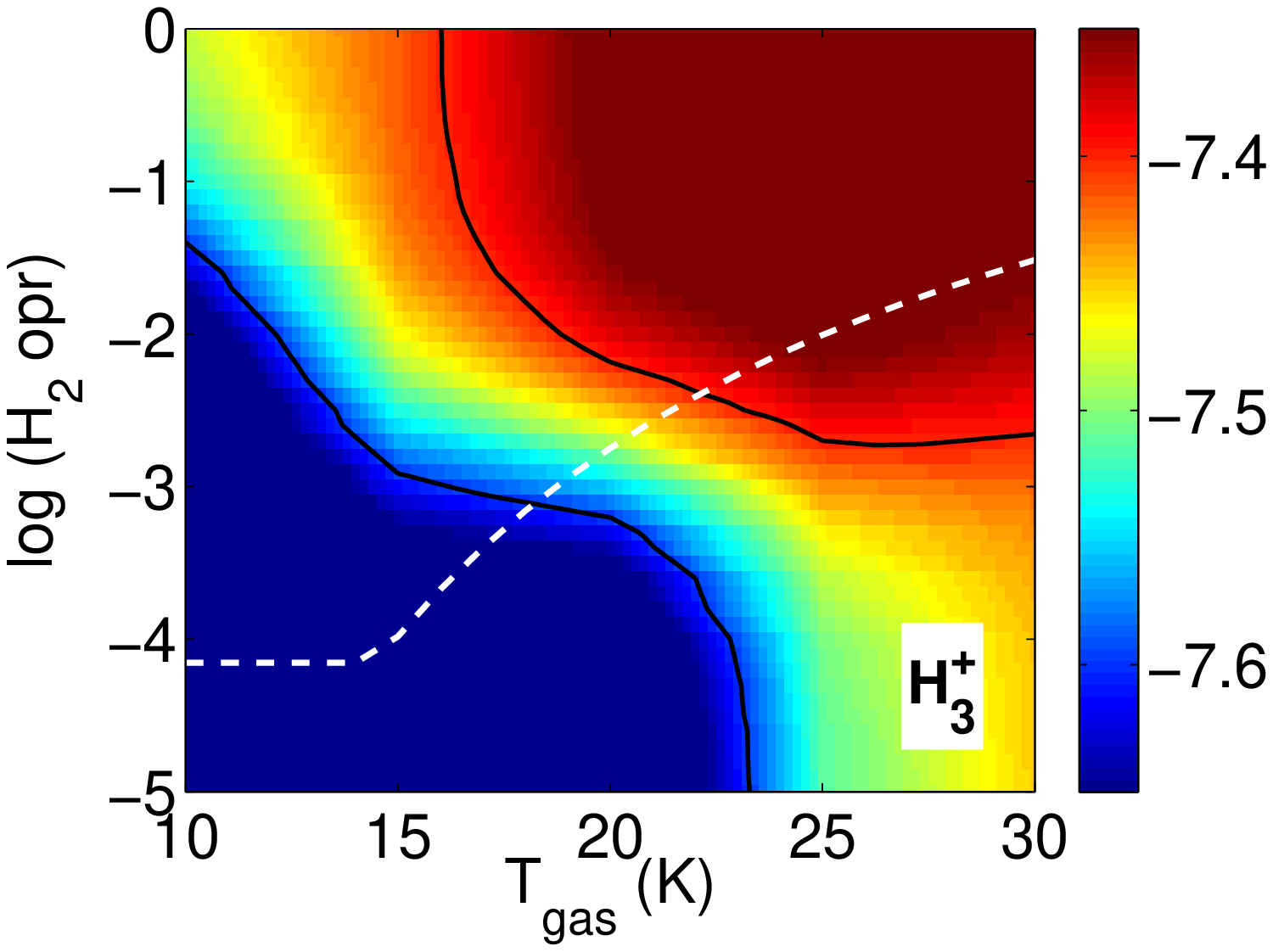}{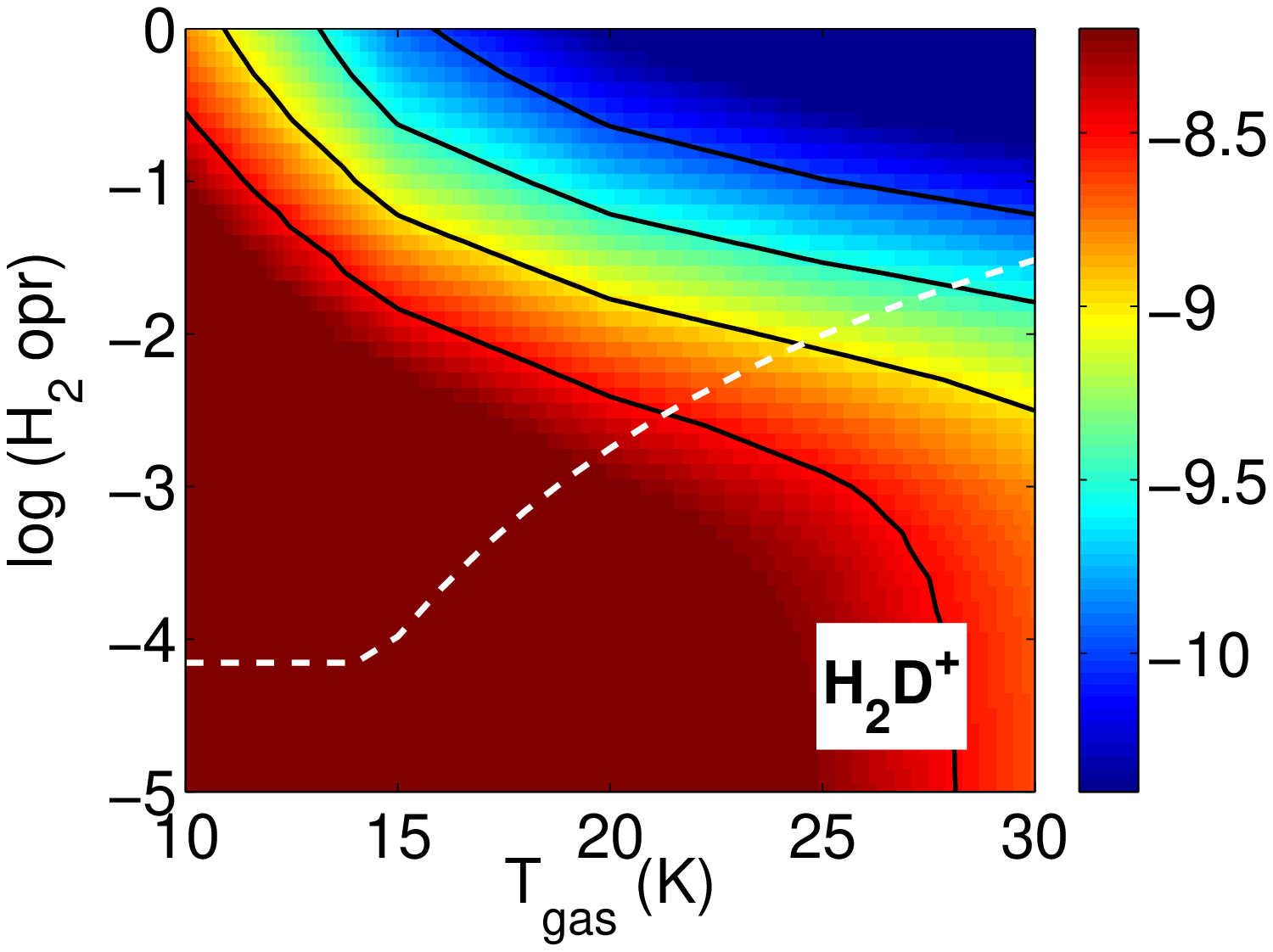}\\
\plottwo{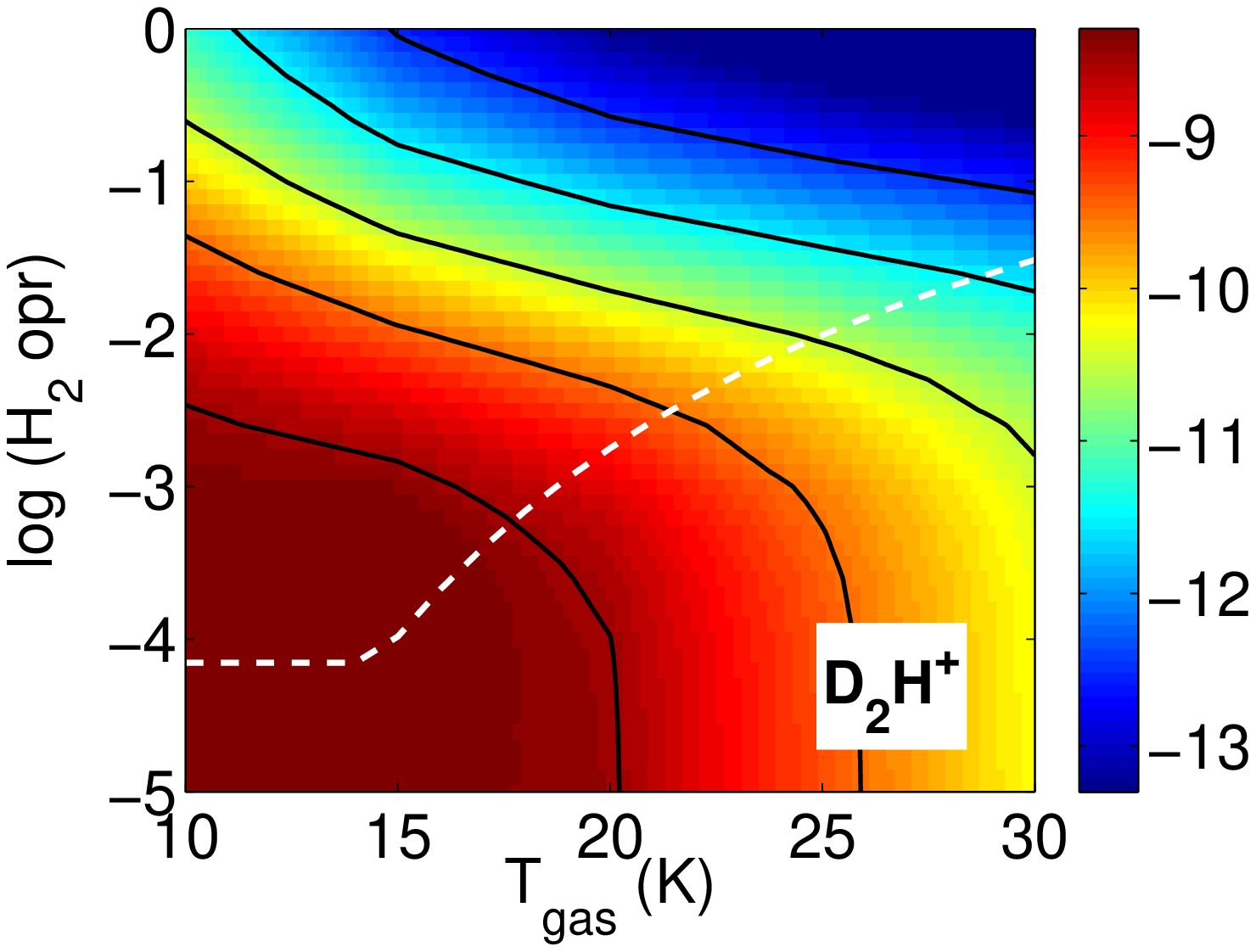}{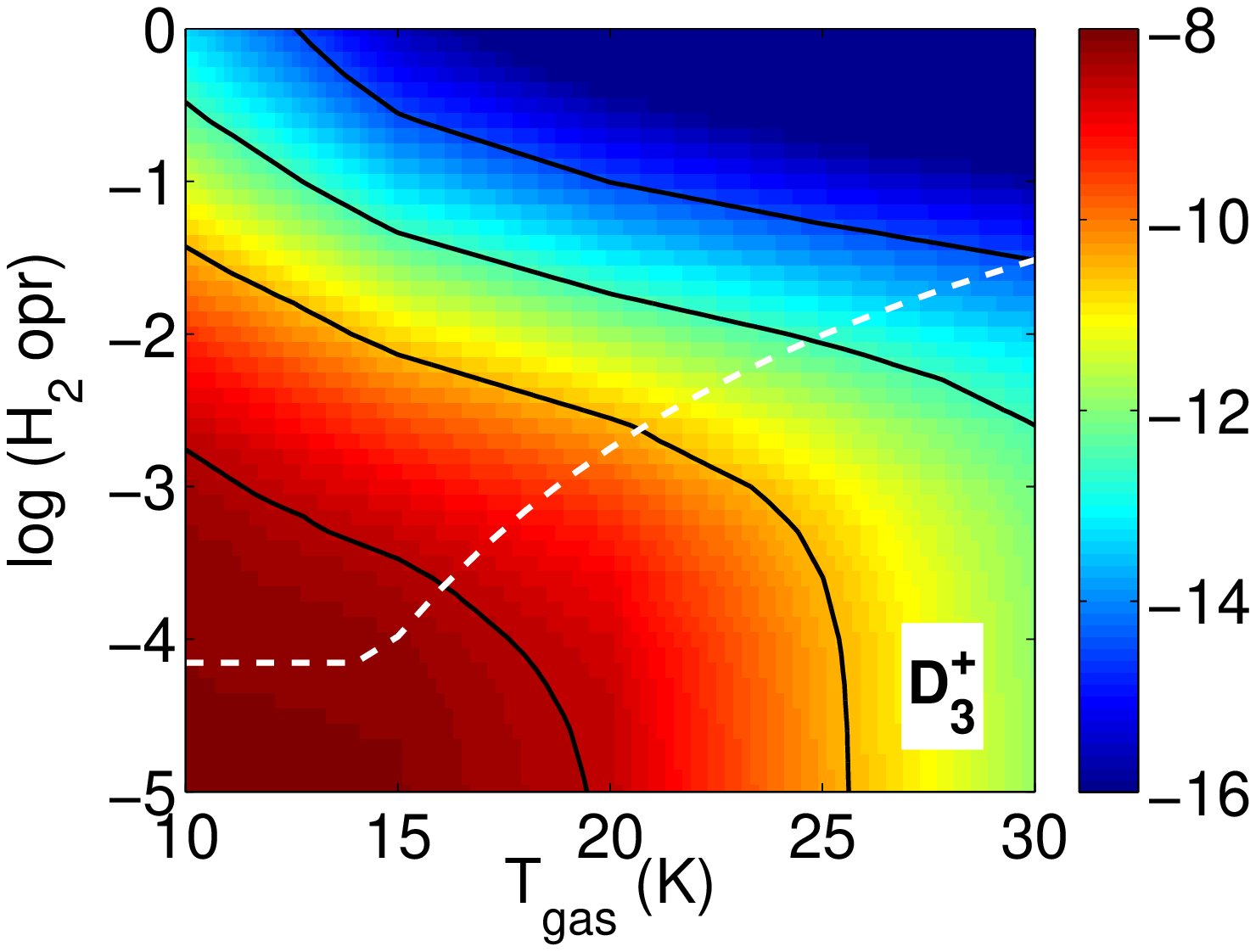}
\figcaption{Abundances of \Hthreep\ and its deuterated isotopologues in the same conditions as in Figure 6. The white dashed line indicates the equilibrium \Htwoopr\ at T $>  \sim15$ K, as in Figure 6. The contours represent $log({\rm X})$ of -7.4 and -7.6 for \Hthreep, -8.5, -9.0, -9.5, and -10 for \HtwoDp, -8.5, -9.5, -10.5, -11.5, and -12.5 for \DtwoHp, and -8.5, -10.5, -12.5 and -14.5 for \Dthreep, respectively, from the red to blue colors. The variation of X(H$_3^+$) is small compared to other deuterated species. This figure can provide a template for \Htwoopr. The observations of \Hthreep\ and its isotopologues can constrain the \Htwoopr\ if the physical conditions of a core are known based on other observations such as submillimeter continuum or
NH$_3$ inversion lines.
}
\end{figure}

\begin{figure}
\figurenum{8}
\epsscale{0.7}
\plotone{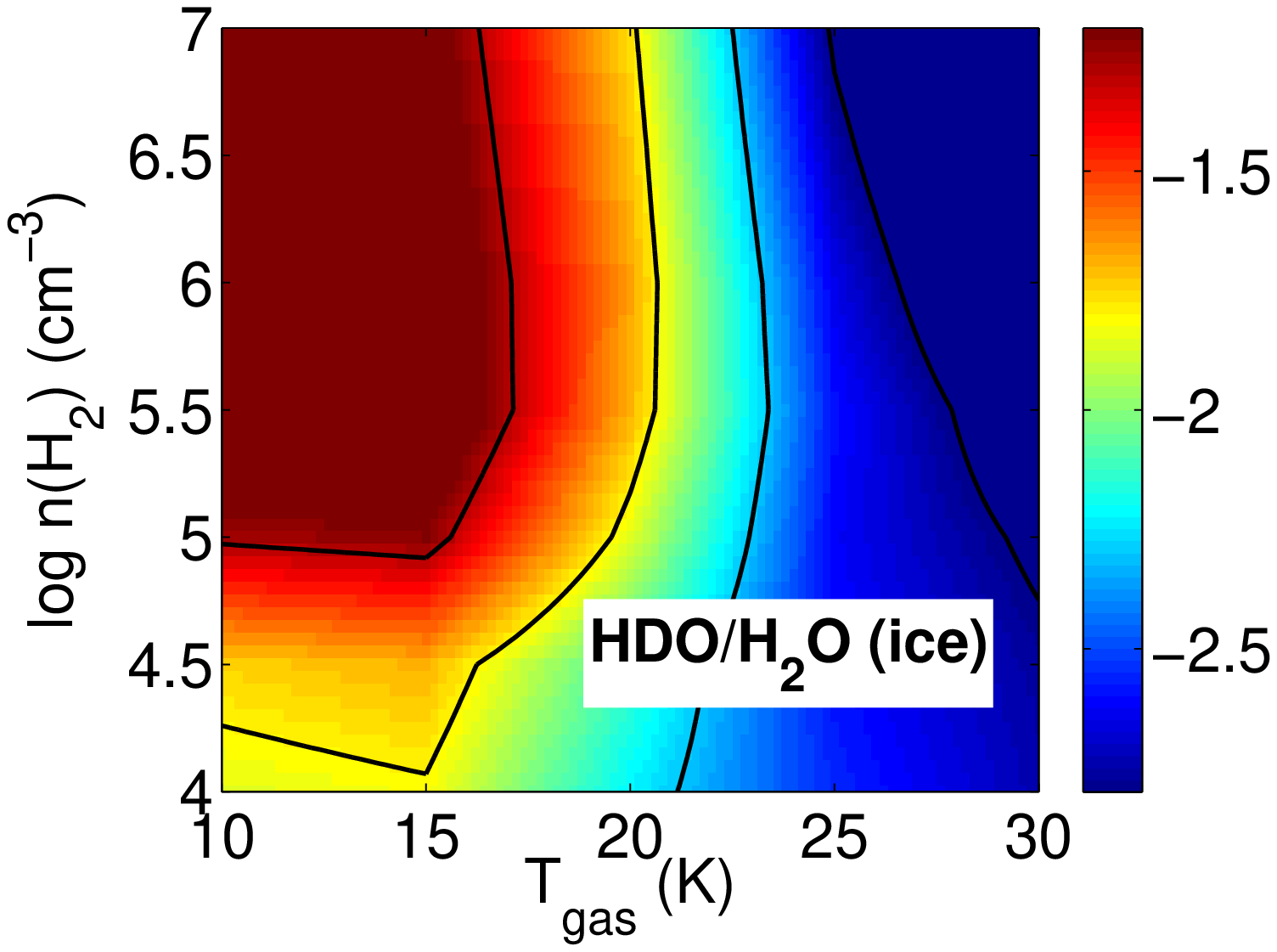}
\plotone{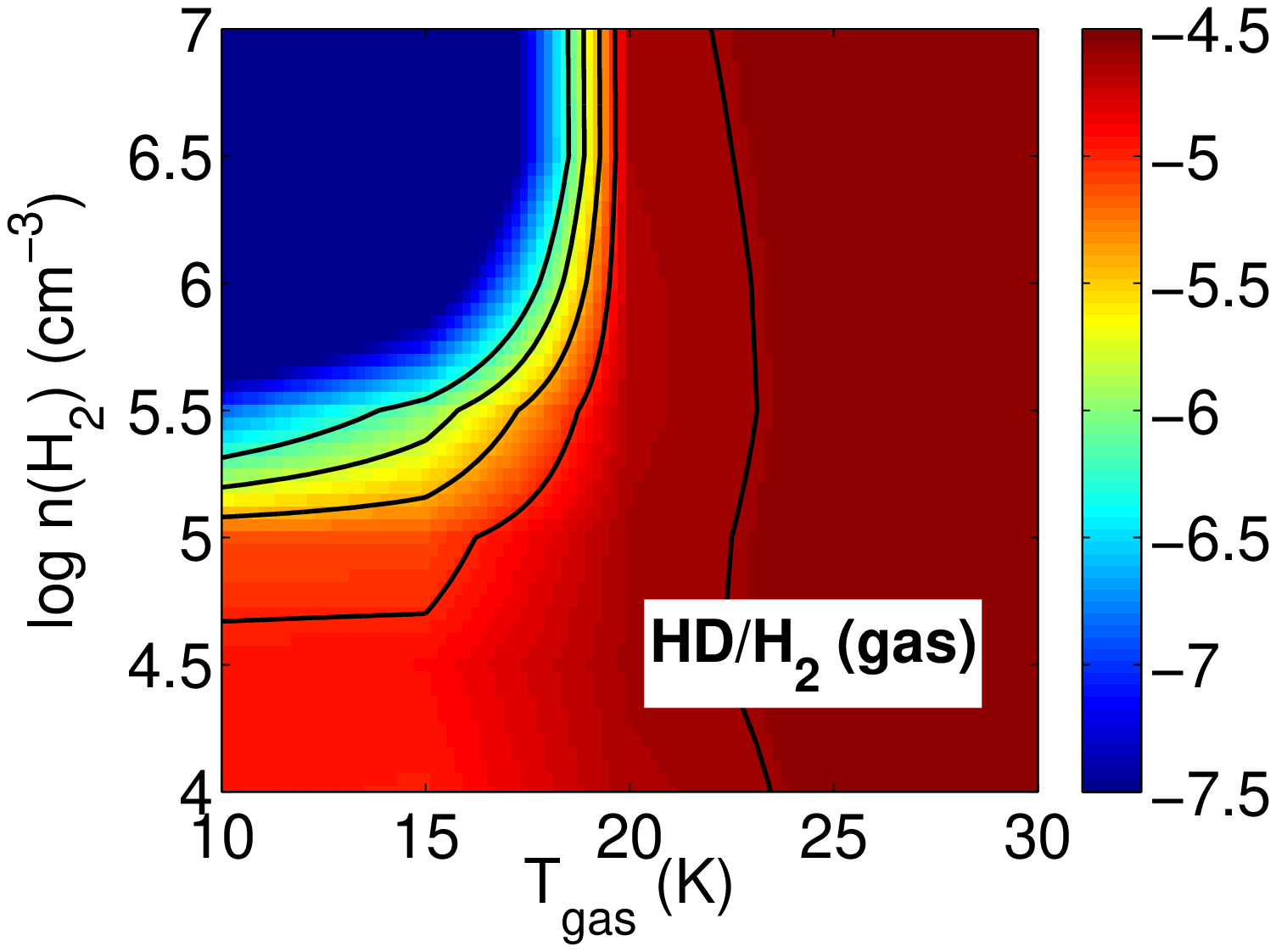}
\figcaption{The D/H ratio ($log({\rm D/H})$) of water ice (top) and molecular hydrogen gas (bottom) in the two dimensional grid of the gas temperature and gas density at the timescale of $10^7$ yrs. The dust temperature is the same as the gas temperature in our models. The fiducial initial abundances were adopted for the calculation. The contours represent $log({\rm D/H})$ of -1.3, -1.8, -2.3 and -2.8 for water ice and  -4.6, -5.0, -5.4, -5.8 and -6.2 for molecular hydrogen gas from the red to blue colors. The equilibrium \Htwoopr\ at each temperature, which is shown in Figure 6 and 7 as the white dashed lines, has been assumed.
}
\end{figure}

\begin{figure}
\figurenum{9}
\epsscale{1.0}
\plottwo{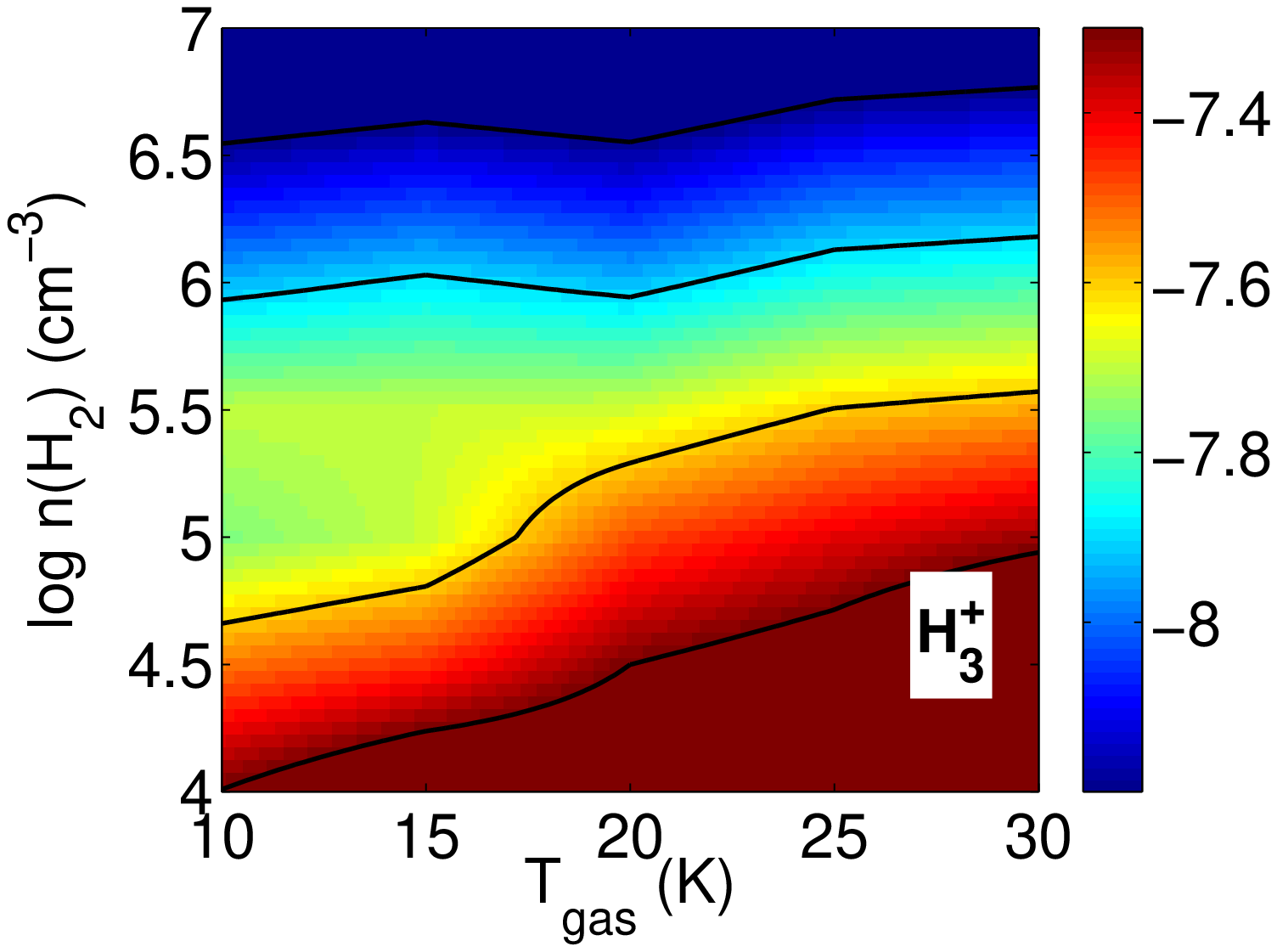}{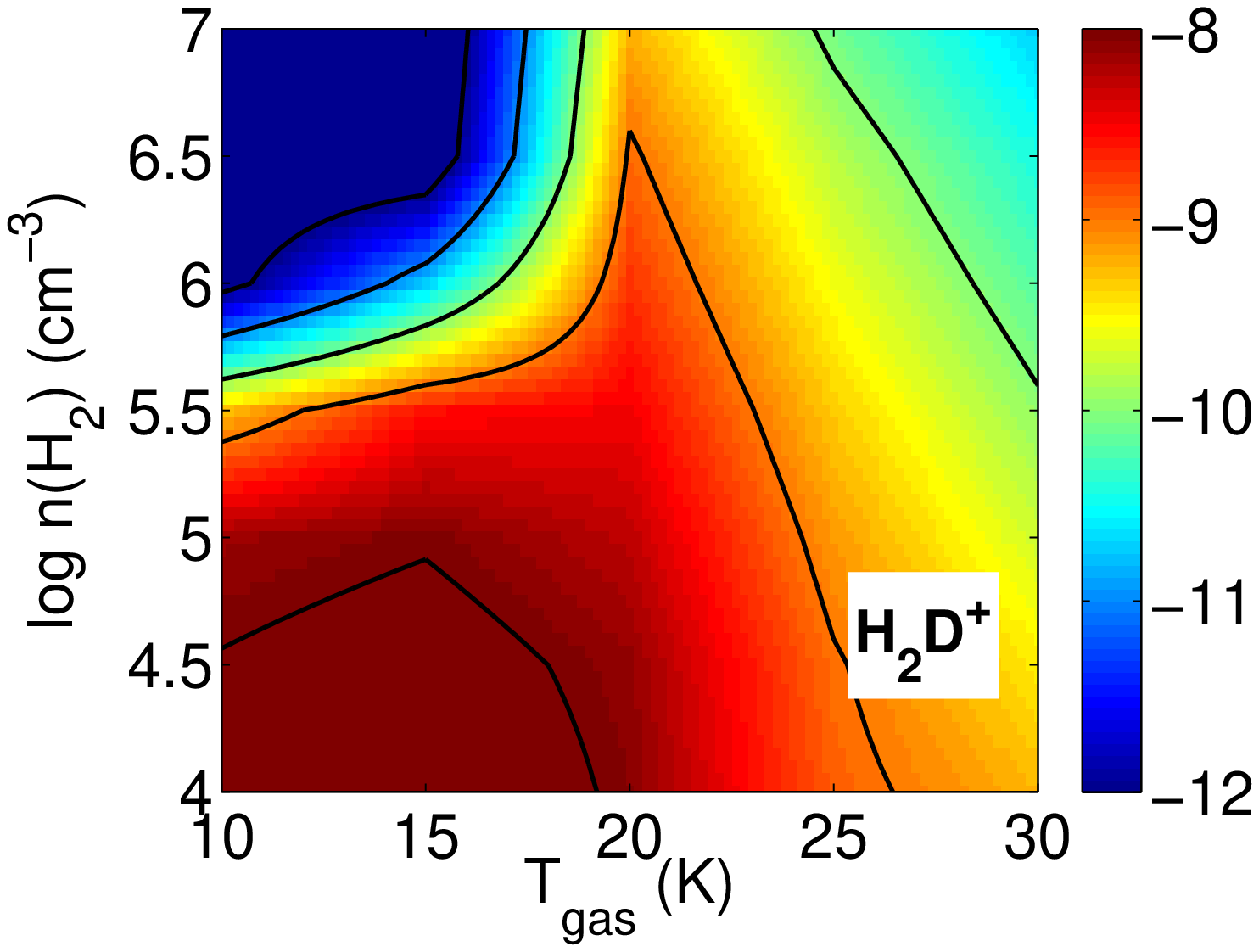}\\
\plottwo{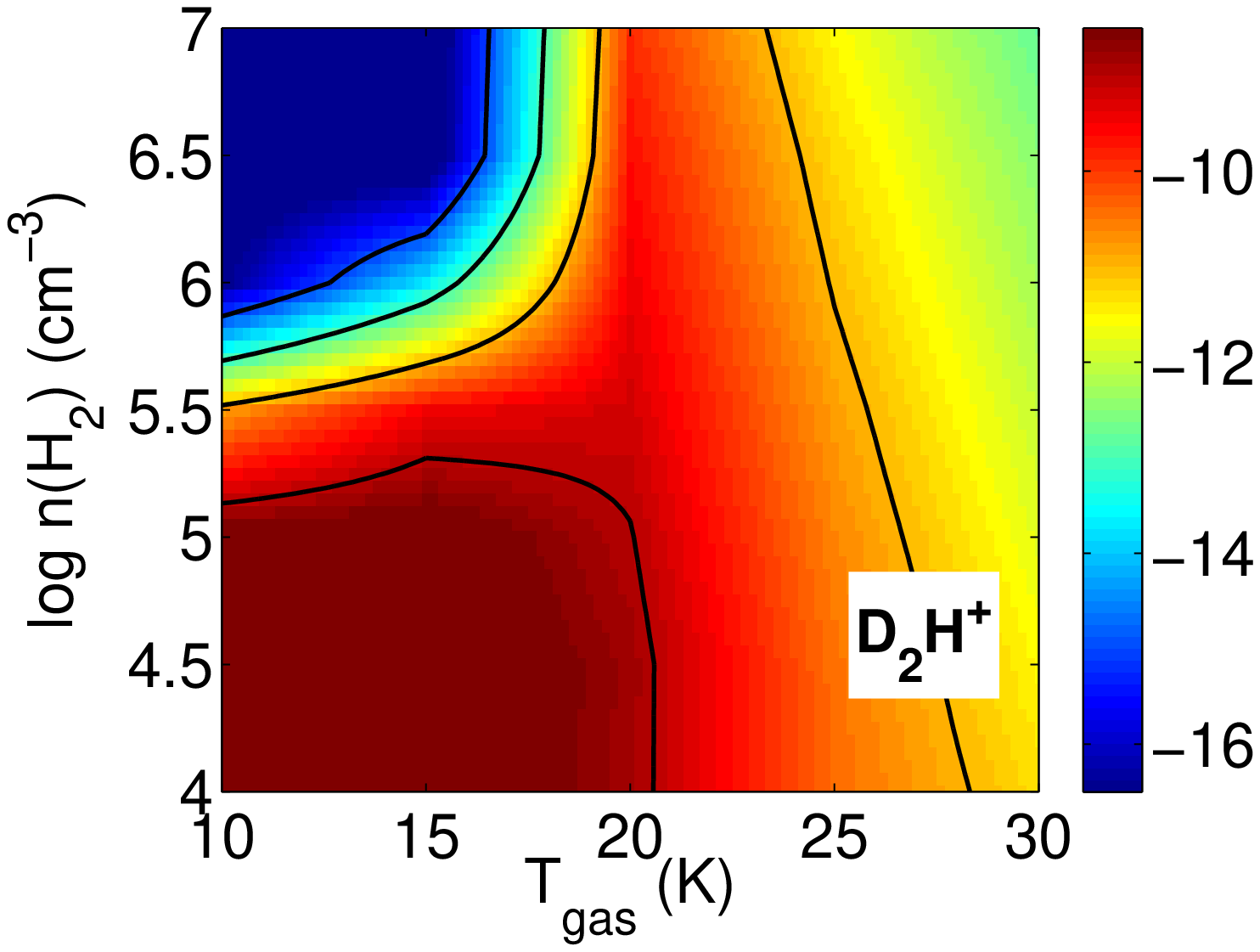}{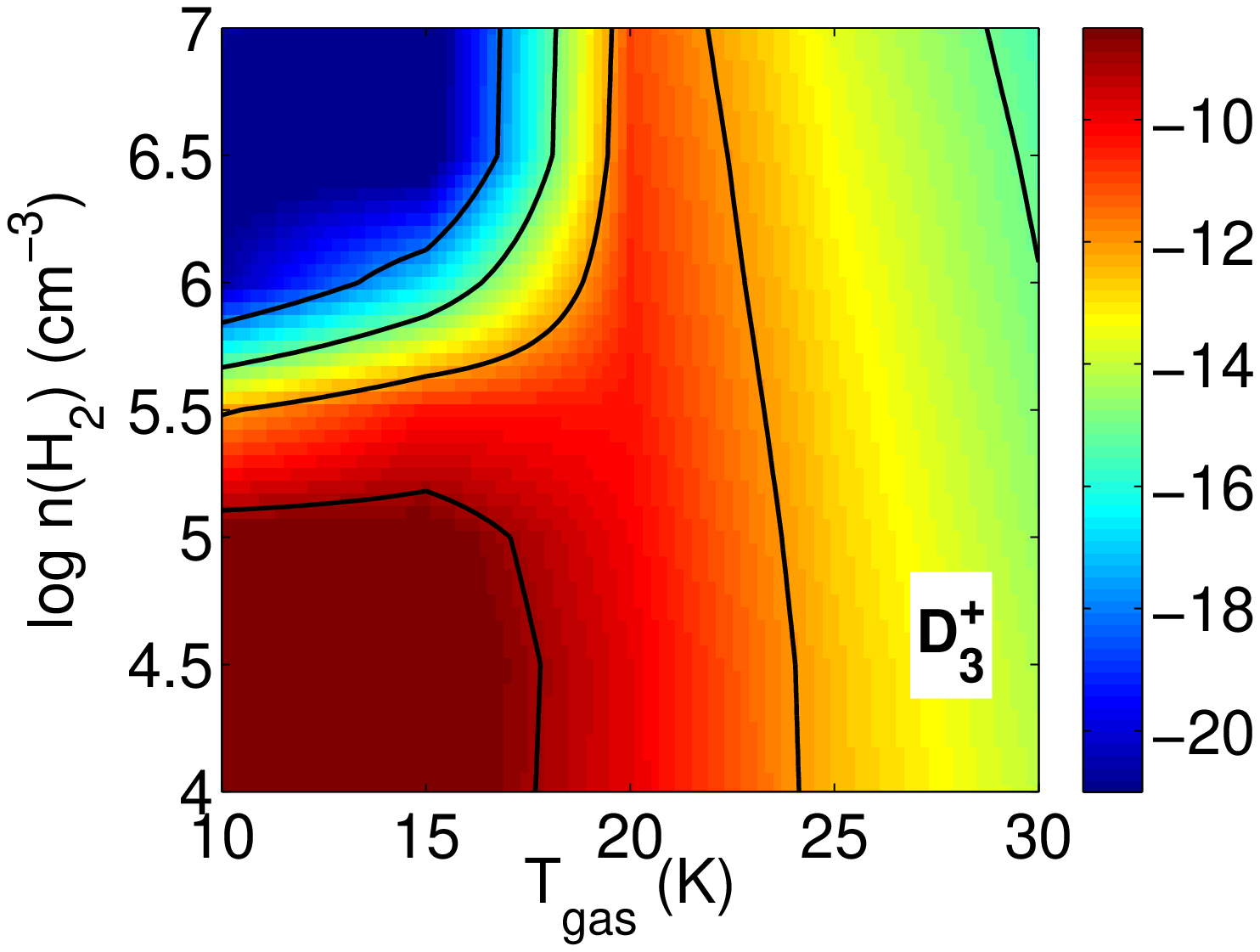}
\figcaption{Abundances of \Hthreep\ and its deuterated isotopologues in the same conditions as in Figure 8. The contours represent $log({\rm X})$ of -7.3, -7.6, -7.9 and -8.2 for \Hthreep, -8.0, -9.0, -10.0, -11.0 and -12.0 for \HtwoDp, -9.0, -11.0, -13.0 and -15.0 for \DtwoHp, and -9.0, -12.0, -15.0, and -18.0 for \Dthreep, respectively, from the red to blue colors. X(H$_3^+$) is almost constant at given temperatures. 
Deuterated species become insensitive to density if T$>$20 K. In the parameter space of the density greater than $3\times 10^5$ and the temperature lower than 20 K, those deuterated species are depleted significantly from gas.
}
\end{figure}

\begin{figure}
\figurenum{10}
\epsscale{0.7}
\plotone{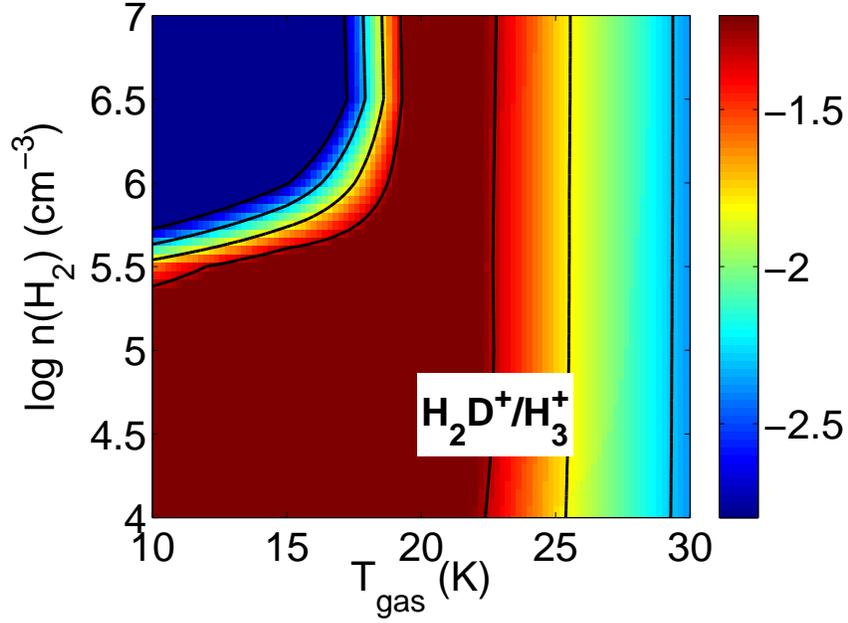}
\figcaption{The D/H ratio ($log({\rm D/H})$) of H$_3^+$ derived from Fig. 10a and 10b. The contours represent $log({\rm D/H})$ of -1.3, -1.8, -2.3 and -2.8, which are the same as used in Fig. 9a.
The D/H ratio of H$_3^+$ becomes very low at the density greater than $3\times 10^5$ \cc\ and the temperature lower than 20 K although it is still higher than the cosmic value. }
\end{figure}

\end{document}